\documentclass{aastex631}
\usepackage{natbib}
\begin{document}

\newcommand{\coj}{CO J=3--2}
\newcommand{\co}{CO}
\newcommand{\kms}{$\mathrm{km\,s^{-1}}$}

\shorttitle{ALMA Survey of Upper Sco}
\shortauthors{Carpenter et al.}

\title{Extending the ALMA Census of Circumstellar Disks in the Upper Scorpius OB Association}

\author[0000-0003-2251-0602]{John M. Carpenter}
\affiliation{Joint ALMA Observatory, Avenida Alonso de C\'ordova 3107, Vitacura, Santiago, Chile}

\author[0000-0001-9223-9091]{Taran L. Esplin}
\affiliation{Steward Observatory, University of Arizona, 933 North Cherry 
Avenue, Tucson, AZ 85721, USA}
\altaffiliation{Strittmatter Fellow}

\author[0000-0003-2822-2951]{Kevin L. Luhman}
\affiliation{Department of Astronomy and Astrophysics,
The Pennsylvania State University, University Park, PA 16802, USA}
\affiliation{Center for Exoplanets and Habitable Worlds, The
Pennsylvania State University, University Park, PA 16802, USA}

\author[0000-0003-2008-1488]{Eric E. Mamajek}
\affiliation{Jet Propulsion Laboratory, California Institute of Technology, 4800 Oak Grove Drive, Pasadena, CA 91109, USA}
\affiliation{Department of Physics and Astronomy, University of Rochester, P.O. Box 270171, 500 Wilson Boulevard, Rochester, NY 14627-0171, USA}

\author[0000-0003-2253-2270]{Sean M. Andrews}
\affiliation{Center for Astrophysics \textbar\ Harvard \& Smithsonian, 60 Garden Street, Cambridge, MA 02138, USA}

\correspondingauthor{John M. Carpenter}
\email{john.carpenter@alma.cl}

\begin{abstract}

We present ALMA Band 7 continuum (340~GHz) and \coj\/ observations for an extended sample of disks in the Upper Scorpius OB Association (Upper Sco, age $\sim$ 10 Myr). The targets were selected from previous studies that identified new members of Upper Sco using photometry and astrometry from the {\it Gaia} mission, and the presence of a disk has been inferred from mid-infrared excess emission. The new ALMA observations are combined with previous ALMA data to define a sample of 202 Upper Sco members with disks that have spectral types between G0 and M5.5. Among these sources, 120 (59\%) have been detected in the continuum with a signal-to-noise ratio $\geq$ 3, and 83 (41\%) have been detected in \coj. Both the continuum and \coj\/ fluxes show a strong correlation with the spectral type of the central star and the type of disk inferred from the shape of the infrared spectral energy distribution, where disks around earlier type stars and full disks are more luminous than disks around later type stars and evolved and debris disks. The median dust continuum luminosity is lower for disks in Upper Sco than in younger regions, as found in previous studies, where the differences are more pronounced in later spectral types (M4--M5) than in earlier spectral types. 
\end{abstract}

\section{Introduction}

Demographic studies are a widely used method to investigate the evolution of protoplanetary disks. By observing the properties of disks surrounding stars of various ages and masses, any age-related variations are assumed to indicate secular evolution. In theory, changes in disk size and mass over time can help constrain the mechanisms driving disk dispersal (see review by \citealt{Manara23}).

Numerous demographic studies have been performed over a range of wavelengths to trace various aspects of disk evolution. Near-infrared (1-4 $\mu$m) photometric surveys, which probe warm dust in close proximity to the star, have shown $\gtrsim 80\%$ of solar-mass stars contain an inner disk at an age of $\sim$1~Myr, and that by an age of $\sim$5~Myr, only $\sim$20\% of the inner disks remain \citep{Haisch01,Hernandez05,Mamajek09}. The {\it Spitzer Space Telescope} and the all-sky {\it Wide-field Infrared Survey Explorer} ({\it WISE}) have extended these studies to mid-infrared wavelengths (3.5--24$\mu$m), which are particularly sensitive to the presence of dust emission at a few to several astronomical units around solar-type stars. \citet[see also \citealt{Ribas14}]{Ribas15} compiled observations from numerous studies with {\it Spitzer} and {\it WISE}. Averaged across all stellar masses, they found an exponential dissipation timescale of $\sim$ 3~Myr at 3.4--12$\mu$m and $\sim$5~Myr at 22--24$\micron$, with shorter dissipation timescales for stars more massive than 2~M$_\odot$. The longer dissipation timescales at long wavelengths are consistent with inside-out clearing of disks (see, however, \citealt{Maeshima21}).

Measurements of the gas dissipation times are more limited but suggest similar timescales. The average mass accretion rate decreases with stellar age \citep{Muzerolle00,Sicilia10,har16}, although individual disks with high accretion rates can persist to ages of $\sim$10~Myr \citep{Ingleby14,Manara20}. \citet{Fedele10} showed that the fraction of K0--M5 stars with measurable gas accretion rates has an exponential dissipation timescale of $\sim$ 2.3~Myr, which is comparable if not slightly shorter than the dust dissipation time. \citet{Briceno19} inferred a similar timescale of $2.1\pm0.5$~Myr based on the accretion fraction of K and M-type stars for various stellar population in Orion.

Millimeter and submillimeter wavelength continuum and spectral line observations provide complementary information on the disk demographics. The longer wavelength observations have lower optical depths than infrared data, providing a better measure of the dust content of millimeter-sized grains. In addition, the higher angular resolution from interferometric observations provide better constraints on the spatial extent of disks. In principle, such studies can discern the dominant mechanisms that control the evolution of disks \citep[e.g., see][]{Rosotti19,Zagaria22,Manara23}.

ALMA has greatly expanded the scope of millimeter and submillimeter studies of disks due to vastly improved sensitivity and angular resolution (for a recent review, see \citealt{Manara23}). \citet[see also \citealt{Carpenter14}]{Barenfeld16} showed that the disks in the Upper Sco OB Association (Upper Sco, $\sim$10 Myr) have lower dust masses than the younger Taurus region. This trend with age has been confirmed with observations of other regions, including Lupus \citep{Ansdell16}, Chamaeleon I \citep{Pascucci16}, Ophiuchus \citep{Williams19}, $\sigma$ Orionis \citep{Ansdell17},  IC~348 \citep{Ruiz18}, the Orion Nebula Cluster \citep{Eisner18}, the Orion A molecular cloud \citep{Grant21,vanTerwisga22}, and the eastern portion of NGC 2024 \citep{vanTerwisga20}. $\lambda$ Orionis, which has a comparable age to Upper Sco, also tends to have fainter disks than younger regions \citep{Ansdell20}, supporting the scenario that the differences in the dust masses may be a result of evolution. However, variations in the radiation field across different regions might also influence disk luminosities.  \citet{Trapman20b} estimated that the ambient radiation field in Upper Sco ranges from 10 to 300~$\mathrm{G_0}$, which is significantly stronger than that in the Taurus molecular cloud \citep{Flagey09} and Lupus \citep{Cleeves16}, though weaker than the levels found in the Orion Nebula Cluster \citep[up to $10^7~\mathrm{G_0}$;][]{Aru24} and $\sigma$ Orionis \citep[up to $\sim10^5~\mathrm{G_0}$;][]{Mauco23}. \citet{Trapman20b} showed that the intermediate radiation fields observed in Upper Sco can result in reduced dust masses.

A key region in demographic studies has been Upper Sco. It is one of three subgroups of the Scorpius-Centaurus OB Association (Sco-Cen), which also includes Upper Centaurus-Lupus (UCL) and Lower Centaurus-Crux (LCC). Upper Sco is the youngest of three subgroups. The age of the association is typically considered to be between 5 and 11 Myr. The ages of the K- and M-type stars in Upper Sco inferred from pre-main-sequence evolution tracks averages to $\sim$ 5~Myr \citep{Preibisch02,Slesnick08}. However, \citet{Pecaut12} inferred an age of $\sim$ 11~Myr for the association based on the luminosities of the F-type stars, the kinematic age expansion of Upper Sco, and isochronal ages for the B-, A-, and G-type stars and the M supergiant Antares. \citet{Feiden16} suggests that the  apparent discrepancy  in the ages between the early and late-type stars is a result of magnetic inhibition of convection of late-type stars, and showed models that include magnetic fields \citep{Feiden12,Feiden13} yield older ages (9--10~Myr) for the late type stars that are more inline with early type stars. More recent studies incorporating {\it Gaia} astrometry suggest that there are distinct clusters within the association that have ages between 4 and 19~Myr \citep{Kerr21,Ratzenbock23_cluster,Ratzenbock23_time}. Regardless of the precise ages and complex star formation history, Upper Sco is generally considered older than regions such as Taurus and Ophiuchus on the basis of the relative luminosity of stars of similar spectral types \citep{Pecaut12,Luhman20b}. And given the mean distance of $\sim$ 145~pc \citep{deZeeuw99,Wright18,Kerr21,Ratzenbock23_cluster}, Upper Sco contains the largest nearby population of disks at its age \citep{Esplin18,luh22disks}.

\citealt{Carpenter14}, \citet{Barenfeld16}, and \citet{Barenfeld17} provided an initial census of the gas and dust properties for a sample of 106 stars in Upper Sco between spectral types of G0 and M5. Since those surveys, the census of the Upper Sco membership has expanded and refined with additional ground-based photometric and spectroscopic surveys and with high-precision astrometry from the {\it Gaia} mission
\citep{per01,deb12,gaia16b}. Given the significance of Upper Sco in demographic studies, we have expanded the submillimeter census of the disk properties in Upper Sco using ALMA. This study present the observation data and re-evaluates the demographic trends. Moreover, the enlarged sample enables us to analyze subsets of the data with spectral type.

The paper is organized as follows. Section~\ref{sample} describes how the sample of stars was selected. Section~\ref{obs} provides an overview of the new observations. Sections~\ref{continuum} and \ref{gas} present the continuum and spectral line measurements, respectively, along with an analysis of the Upper Sco demographics. The results and a comparison with other regions are discussed in Section~\ref{discussion}.

\section{Sample}
\label{sample}

The sample for this survey was drawn from the census of young stars and brown dwarfs presented in \citet{Luhman18b}, who identified 1631 likely members of the Upper Sco association based on parallaxes, proper motions, color-magnitude diagrams, and optical and infrared spectra. \citet{Esplin18} identified stars with circumstellar disks by mid-infrared excess emission detected with the {\it WISE} and {\it Spitzer}. From the \citet{Esplin18} compilation, 284 disk-bearing stars were selected that have spectral types between G0 and M5. (In a later analysis, one of the M5 stars was reclassified as M5.5.) The sources were observed with ALMA as described in Section~\ref{obs}.

Since the analysis by \citet{Luhman18b} and \citet{Esplin18}, the second and third data releases of {\it Gaia} have become available \citep[DR2 and DR3,][]{GaiaDR2,GaiaEDR3,GaiaDR3}, which provide high-precision proper motions and parallaxes down to much fainter magnitudes than the first data release ($G\sim 20$). \citet{Luhman20b} and \citet{Luhman22} used the data from DR2 and DR3 to refine the census of Upper Sco and other populations in Sco-Cen. Because of overlap among the Sco-Cen populations in their spatial positions and kinematics, some stars have kinematics that are consistent with multiple groups.

The 284 ALMA targets are listed in Table~\ref{tbl:sample}, which includes each source's designation from {\it Gaia} DR3, adopted spectral type \citep{Luhman20b,Luhman22,Manara20},  disk classification \citep{Esplin18,luh22disks}, parallactic distance from {\it Gaia} DR3 \citep{BailerJones21}, and the Sco-Cen populations with which the {\it Gaia} kinematics are consistent \citep{Luhman22}. The spectral types are taken primarily from the compilation in \citet{Luhman22}. \citet{Manara20} presented spectral types from X-shooter spectra for 36 stars.\footnote{Of the 36 stars, 35 are in our sample. \citet{Manara20} resolved 2MASS J15354856-2958551 into a binary for the first time and the secondary is not explicitly part of the sample.}
For five stars where the spectral type in \citet{Manara20} differed from \citet{Luhman22} by more than one subclass, we adopted the classification from \citet{Manara20}. If {\it Gaia} data are not available or the star is not a kinematic member of a known Sco-Cen subgroup, no population is listed in Table~\ref{tbl:sample}. Among the 284 targets, 116 have kinematics that are consistent with Upper Sco alone, 92 are consistent with Upper Sco and one or more other Sco-Cen groups, 76 are not members of Upper Sco or lack the kinematic data from {\it Gaia} DR3 for membership classifications. For this study, we adopt the 208 stars from the first two categories as members of Upper Sco. As described in Section~\ref{subsec:flux}, for 13 out of the 284 sources, no continuum emission was detected around the target star but a continuum source was detected nearby that could be responsible for the observed infrared excess. These 13 sources, which include six Upper Sco members, are not considered when analyzing the continuum and CO flux distributions. The 202 sources that constitute the final Upper Sco sample are marked in Table~\ref{tbl:sample}. 

\begin{figure}
\centering
\includegraphics[width=\columnwidth]{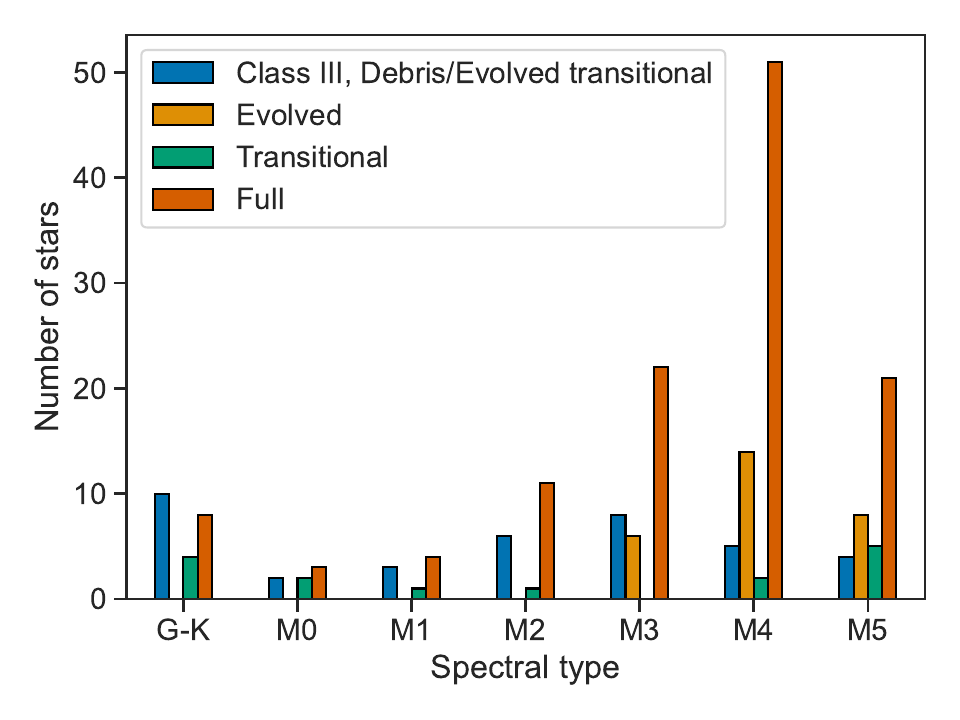}
\caption{Distribution of spectral types for the 202 sources in the Upper Sco sample. One source (2MASS J16193570-1950426) with a uncertain spectral type has been excluded. For the M-type stars, the plotted subclass is truncated at the integer value (e.g, M2.5 is plotted as M2).}
\label{fig:sample}
\end{figure}

As described in \citet{Esplin18} and \citet{luh22disks}, the disks have been classified based on the shape of the spectral energy distribution and location in infrared color-color diagrams. The final Upper Sco sample includes 121 sources where the disk has been classified as ``full'' (where the disks is optically thick at infrared wavelengths and lack significant clearing of dust in the inner disk), 15 as ``transitional'' (where the disks have inner holes of radius $\gtrsim$~1~au), 28 as ``evolved'' (the disk is becoming optically thin but no significant clearing of the inner disk has occurred), and 38 as ``debris'', ``evolved transitional'' or class III, where the disk is optically thin and have inner holes of radius $\gtrsim$~1~au. Figure~\ref{fig:sample} shows the distribution of the spectral types by the disk class. Most stars (72\%) have spectral types of M3 or later, and among the later type stars, 64\% have full disks. 

\section{Observations}
\label{obs}

ALMA observations of the continuum and Carbon Monoxide (CO) were obtained in Cycles 0, 2, and 6 using the 12-m array. The Cycles 0 and 2 data for 106 sources were presented in \citet[see also \citealt{Carpenter14}]{Barenfeld16}. An additional 178 sources were observed in Cycle 6  following the observational setup in the earlier cycles. 

Table~\ref{tbl:obs} presents the observation log for the Cycle 6 observations. An observation log for the earlier observations is available in \citet{Barenfeld16}. As with the previous surveys, all observations in Cycle 6 were obtained in Band 7 \citep{Band7}. Spectral windows were centered at 334.2, 336.1, 346.2, and 348.1~GHz with a bandwidth of 1.875~GHz for each window. The window centered at 346.2~GHz, which includes the \coj\ transition at a rest frequency of 345.7959899~GHz, has a channel width of 0.488~MHz (0.429~\kms\/ for \coj) with a spectral resolution of 0.977~MHz (0.858~\kms). The other three windows have a channel width and a spectral resolution of 0.977~MHz and 1.129~MHz, respectively.

The observations were obtained with the 12-m array using between 43 and 50 antennas. The maximum baselines ranged between 740~m to 2617~m to provide an angular resolution between 0\farcs1 and 0\farcs3. The full-width-at-half-maximum (FWHM) primary beam size of the observations is 18\farcs5. The typical on-source integration time was 2.5 minutes to provide a continuum sensitivity (1$\sigma$) of $\sim$ 0.15~mJy.

The  data were calibrated by ALMA staff using the ALMA data  reduction pipeline. Data reduction steps include atmospheric calibration using the 183~GHz water vapor radiometers, bandpass calibration, flux calibration, and gain calibration. The calibrators for each observation date are listed in Table~\ref{tbl:obs}. We performed phase-only self-calibration for the bright sources using Common Astronomy Software Applications (CASA) 5.6.1 package \citep{CASA22}. We assume a 1$\sigma$ flux calibration uncertainty of 10\%.

Continuum images were constructed in CASA by combining all channels except for channels with \coj\ velocities (LSRK) between $-$10 and 15~\kms. A \coj\ image cube was generated between velocities of $-$200 and 200~\kms. Images were deconvolved using the CASA task {\tt tclean}. A circular clean mask was centered on the expected stellar position with a radius of 1\arcsec\ for the continuum images and 2\arcsec\ for the \coj\ cubes. A larger radius was used for \coj\/ since the CO emission is often extended over a larger radial extent. Visual inspection of the continuum images identified 40 additional continuum sources that are not associated with the Upper Sco target. A clean mask was also placed around these sources for both the continuum and \coj\ deconvolution. All images were made with robust=2 Briggs weighting to maximize sensitivity.

\begin{figure}
\centering
\includegraphics[width=\columnwidth]{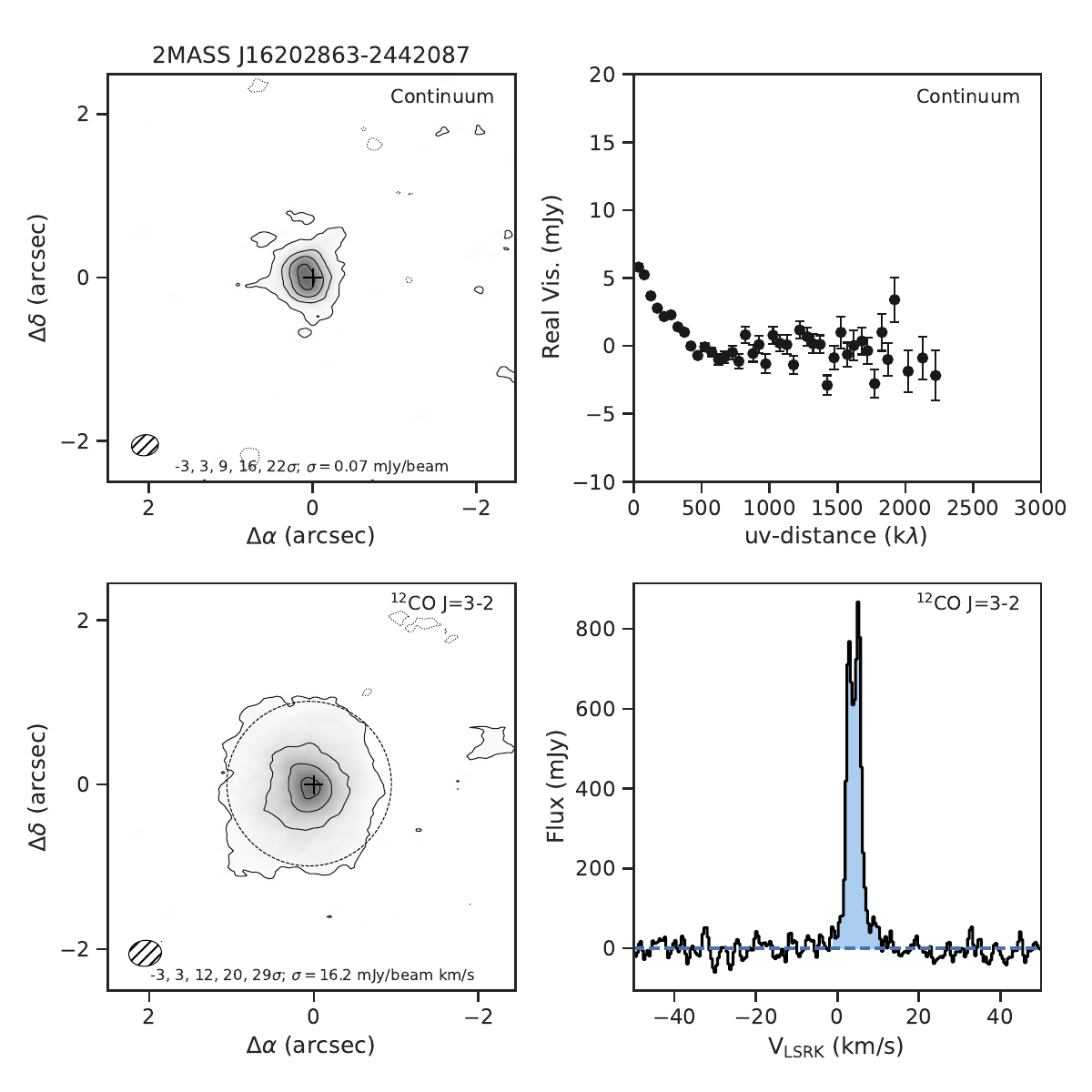}
\caption{ALMA images of 2MASS J16202863-2442087.
{Upper left:} 870~$\mu$m continuum image. 
{Lower left:} Integrated \coj\/ intensity.
{Upper right:} Real part of the visibility versus baseline length, which has been phase shifted to the centroid of continuum emission (if detected) or the stellar position for continuum non-detections.
{Lower right:} Integrated \coj\/ intensity within the circular aperture shown in the lower left panel. If the \coj\/ integrated intensity is detected with a SNR $\geq 3$, the velocity extent used to compute the integrated intensity is shaded in blue if two or more channels have a SNR $\geq 3 \sigma$, and shaded in yellow otherwise.
In the panels on the left, the plus symbol indicates the expected stellar position, the hatched ellipse is the full-width-at-half-maximum of the clean beam, and the dashed circle shows the aperture used to compute the \coj\/ spectrum or the continuum flux if aperture photometry was used. Diamond symbols indicate the location of any additional continuum sources detected in the image. While no primary beam correction has been applied to the images, a primary beam correction has been applied to the spectra.
The complete figure set (284 images) is available in the online journal.
}
\label{fig:images_all}
\end{figure}

Figure~\ref{fig:images_all} presents the continuum and \coj\ data for all 284 sources observed by ALMA. The figure for each source includes the continuum image, the integrated \coj\ intensity image, the real part of the continuum visibility versus the {\it uv} distance, and the \coj\ spectrum. The continuum and CO results are presented in Section~\ref{continuum} and Section~\ref{gas}, respectively.

\figsetstart
\figsetnum{2}
\figsettitle{ALMA images of the complete sample}

\figsetend

\section{Analysis of ALMA Band 7 Continuum Data}
\label{continuum}

\begin{figure*}
\centering
\includegraphics[width=\textwidth]{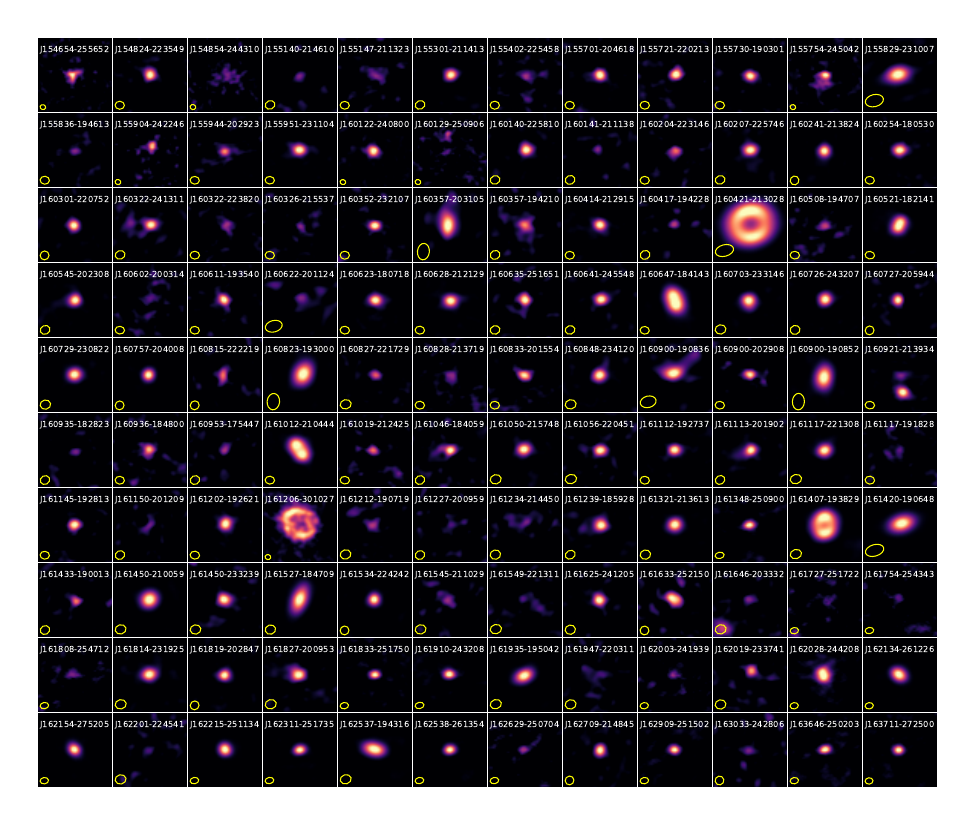}
\caption{Band 7 continuum images for 120 sources that are confirmed Upper Sco members and have been detected in the continuum with a signal to noise ratio $\ge$ 3. The angular size of each image is $3''\times3''$ centered on the expected stellar position. The FWHM synthesized beam is shown in the lower left corner of each panel. The sources are ordered in increasing right ascension from left to right, starting in the upper left corner. 
\label{fig:images_cont_members}}
\end{figure*}

\begin{figure*}
\centering
\plotone{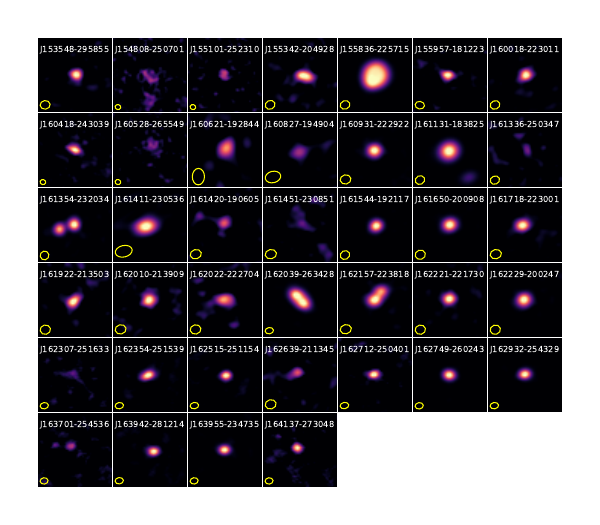}
\caption{Same as Figure~\ref{fig:images_cont_members}, except for the 39 sources in the sample that are not Upper Sco members and have been detected in the continuum with a signal to noise ratio $\ge$ 3. 
\label{fig:images_cont_nonmembers}}
\end{figure*}

Figure~\ref{fig:images_cont_members} shows the Band 7 continuum images for the 120 Upper Sco sources that have been detected with a signal-to-noise ratio of $\ge$ 3 (see Section~\ref{subsec:flux}). For completeness, Figure~\ref{fig:images_cont_nonmembers} presents similar images for the 39 non-members of Upper Sco that were detected with ALMA. Most of the sources are point-like or slightly resolved at the resolution of these observations. Notable exceptions include the dust ring resolved in 2MASS J16042165-2130284 \citep{Mathews12,Zhang14} and 2MASS J16120668-3010270 \citep{Sierra24}. The remainder of this section analyzes the continuum measurements of the Upper Sco sources, including the procedure to measure the fluxes (Section~\ref{subsec:flux}) and analysis of the continuum measurements (Section~\ref{subsec:demographics}).

\subsection{Flux measurements}
\label{subsec:flux}

Fluxes were measured by either fitting the visibility data with a parametric model or performing aperture photometry in the images. First, a preliminary list of sources was generated by visually inspecting the images. If a source contained obvious asymmetries, the flux density was measured in the images using aperture photometry measured with the {\tt imstat} function in CASA. The aperture radius was chosen by visual inspection of the images and analyzing the curve-of-growth using different aperture sizes. The uncertainties in the aperture fluxes were computed from the RMS in the image and the number of independent beams in the aperture. For sources with no obvious substructure, a two dimensional gaussian was fitted to the visibility data. The free parameters in the model are total flux density, the right ascension and declination positional offset from the phase center of the image, the FWHM of the major axis, the ratio of the minor to major axis (which reflects the inclination of the disk), and the position angle. In a few cases, the FWHM of the fitted gaussian to faint sources converged to an unrealistic large value  based on inspection of the images. For these sources, a delta function was fitted with three free parameters: the source flux density and the positional offset. If there was no obvious detection in the images at the stellar position, a point-source model was fitted to the data fixed at the expected stellar position. If a continuum source was detected but it was not coincident with the stellar position, we fitted simultaneously to the visibility data a Gaussian to the continuum source and a delta function at the stellar position, where the latter was used to establish an upper limit to the continuum flux of the Upper Sco source.

The visibility fitting was performed using the affine-invariant Markov chain Monte Carlo (MCMC) ensemble sampler implemented in the Python package Emcee v3.0.2 \citep{Foreman13}. The fits used 50 walkers and the fits were run until the change in the autocorrelation length of the chains varied by less than 1\%. Once convergence was reached, an additional 50,000 steps were run and the distribution of chains for the final 50,000 steps was used to estimate the parameters. Corner plots for the gaussian fits are presented in Appendix~\ref{app:corner}.

The stars with disks were identified by infrared photometric surveys with an angular resolution of $\sim$2-12$''$. Instead of originating from a disk surrounding the target star, the apparent infrared excess could potentially originate from a disk that surrounds a binary companion or from a background source, especially a galaxy. The sub-arcsecond emission with ALMA can be used to determine whether the ALMA detections coincide with the stellar position. Appendix~\ref{appendix:contaminants} compares the observed ALMA positions with the expected stellar positions. Thirteen sources were identified as being contaminated by a nearby source that could potentially be responsible for the observed excess infrared radiation. Although these sources are listed in the flux tables for completeness, they are not used in analyzing the properties of Upper Sco.

Table~\ref{tbl:flux_sample} lists the measured fluxes for the 284 stars in the full sample. The table includes the source name, the Band 7 continuum flux measurement, the positional offset from the expected stellar position, and if applicable, the FWHM of the major axis, the inclination angle, and the positional angle of the Gaussian. Table~\ref{tbl:flux_other} lists the Band 7 continuum fluxes for additional sources detected in the field that are offset from the stellar position.

\subsection{Demographics}
\label{subsec:demographics}

The enlarged sample of Upper Sco members allow us to re-examine the flux distributions with demographics, especially with spectral type of the host star and the disk type inferred from infrared observations.

\subsubsection{Spectral type}
\label{corr_spt}

\begin{figure}
\centering
\includegraphics[width=\columnwidth]{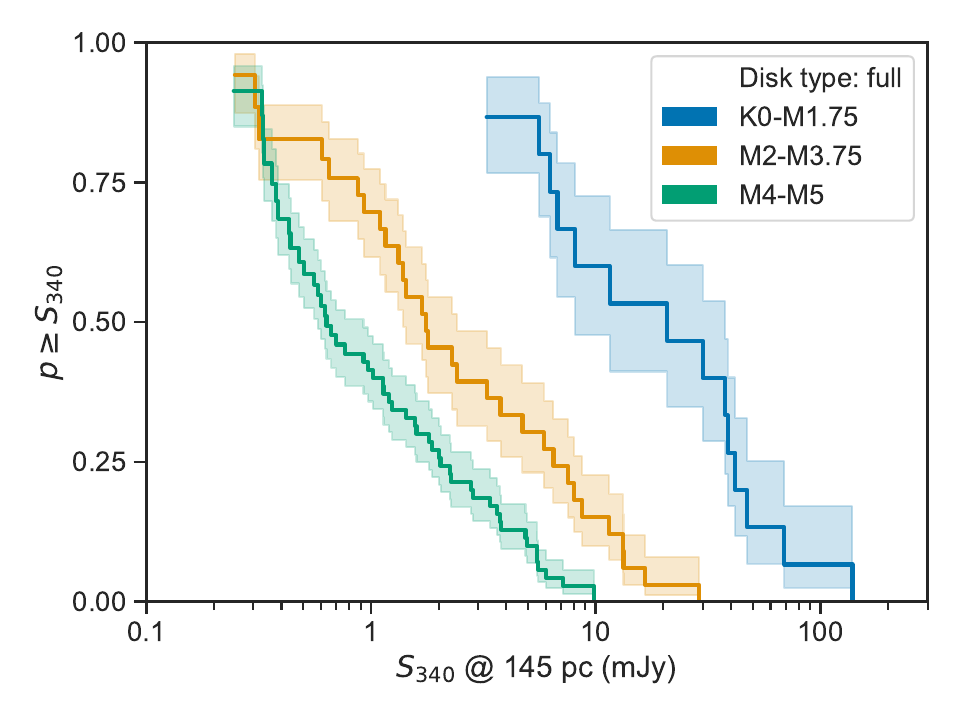}
\caption{The Kaplan-Meier estimator of the Band 7 continuum fluxes for the full disks in Upper Sco, grouped in three spectral type ranges. Fluxes are for the observed frequency of 340~GHz at a distance of 145~pc. The results show that disks around earlier spectral types have more luminous disks than stars with later spectral types. 
\label{fig:groupspt}}
\end{figure}

Figure~\ref{fig:groupspt} shows the Kaplan-Meier estimator (as implemented in {\tt Lifelines}; \citealt{Lifelines}) of the Band 7 continuum fluxes for the Upper Sco sample. As demonstrated in Section~\ref{corr_disktype}, the submillimeter fluxes correlate with the disk type. Therefore, we select only ``full'' disks since it provides the largest subsample. The results show a clear dependence of the submillimeter continuum flux distributions on spectral type. Full disks that surround stars with spectral types K0--M1.75 are clearly brighter than M2--M3.75 stars, which in turn are brighter than the disks around M4--M5 stars. The log-rank two-sample linear rank test as implemented in the {\tt R} \citep{R} package {\tt EnvStats} \citep{EnvStats} confirms these results. The probability that the disks around K0--M1.75 stars have the same luminosity distribution as M2---M3.75 and M4-M5 stars is $3\times10^{-4}$ and $3\times10^{-13}$ respectively. When comparing M2--M3.75 and M4--M5 stars, the probability that the two distributions are drawn from the same parent population is $0.02$.

\subsubsection{Disk type}
\label{corr_disktype}

\citet{Esplin18} classified the disks in Upper Sco based on the infrared colors from 2MASS and {\it WISE}, and in a few cases, the disk classification was updated by 
\citet{Luhman20b} and \citet{luh22disks}. We investigate in this section the dependence of the submillimeter luminosities on the disk type inferred from the mid-infrared colors. 

The  detection rate has a clear dependence on disk type. The full and transitional disks have relatively high detection rates of 80\% (97/121) and 93\% (14/15) respectively. In contrast, evolved disks have a detection rate of 32\% (9/28), and none of the 38 debris-type disks (i.e., debris/evolved transitional or Class III) have been detected. 

These results do not depend on whether the star is an unambiguous kinematic member of Upper Sco or is a possible member (see Section~\ref{sample} and Table~\ref{tbl:sample}). The Band 7 continuum detection rate for full disks for unambiguous Upper Sco members versus possible members is 79\% and 81\% respectively. Similarly, transitional (100\% vs. 86\%), evolved (31\% vs. 33\%), and debris-type disks (0\% vs. 0\%) all have similar detection rates. This suggests that most of the possible Upper Sco members are true members of the association or that the detection rates do not evolve significantly between Upper Sco and the Lower Centaurus Crux and Upper Centaurus Lupus associations.

\begin{figure*}
\centering
\plottwo{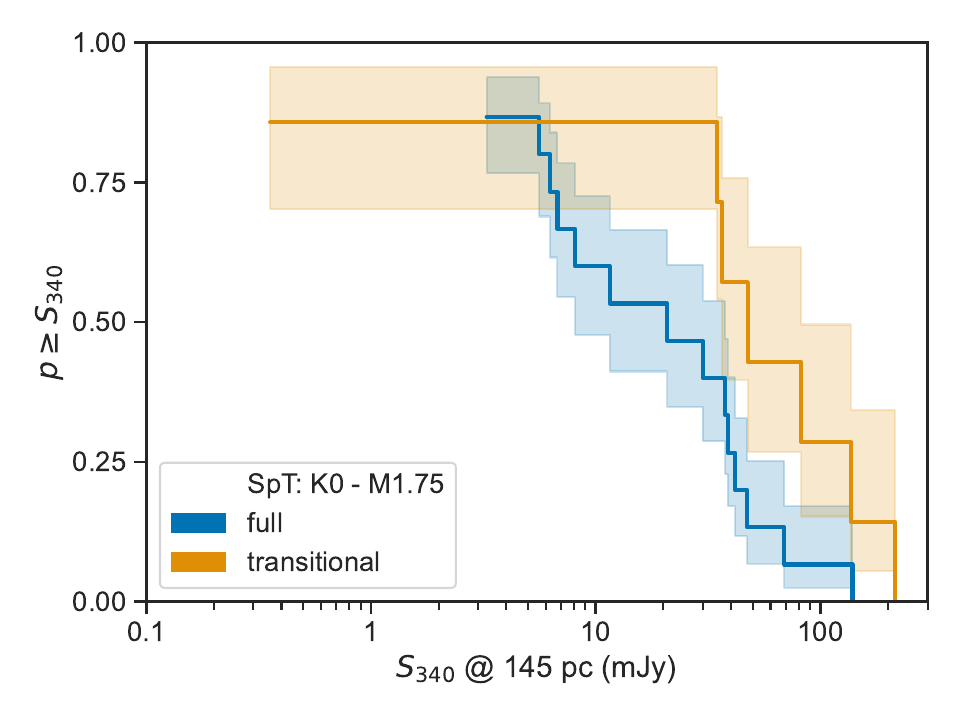}{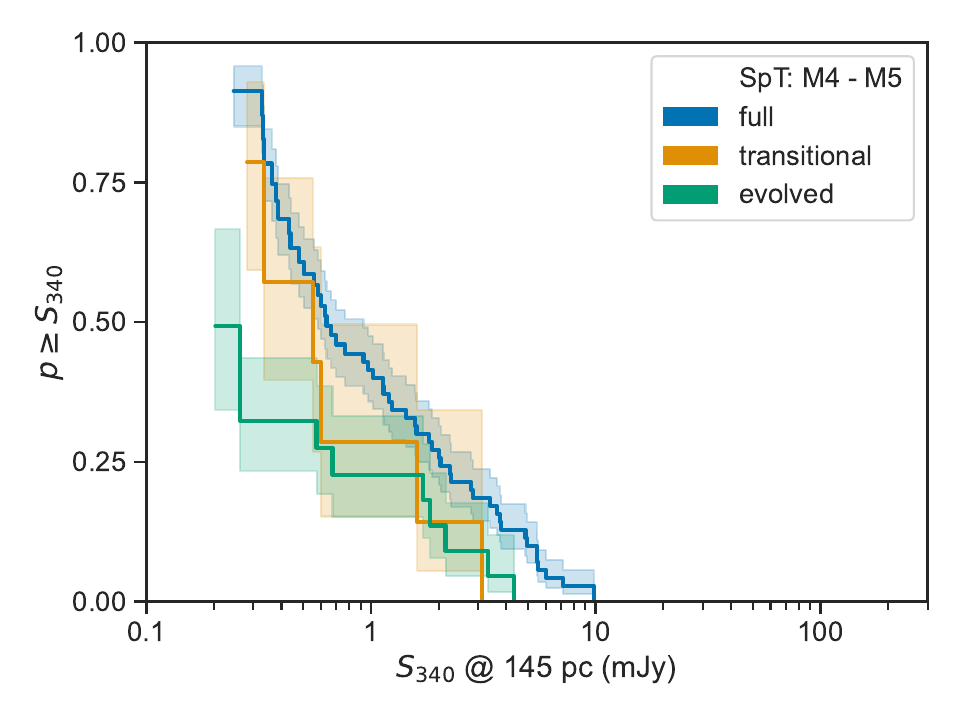}
\caption{The Kaplan-Meier estimator of the Band 7 continuum fluxes for disks in Upper Sco classified for various disk types. The left panel shows the results for stars with spectral type between K0 and M1.75, and the right panel for stars between M4 and M5. No evolved disks are shown in the left panel since none of these systems in this spectral type range were detected with ALMA. All fluxes are for the observed frequency of 340~GHz at a distance of 145~pc. 
\label{fig:grouptype}}
\end{figure*}

Figure~\ref{fig:grouptype} shows the Kaplan-Meier distributions of the submillimeter continuum fluxes by disk types for various spectral-type ranges. The left plot shows the spectral type ranges between K0 and M1.75. Evolved and debris-type disks are not shown since none were detected by ALMA in this spectral type range. While transitional disks are more luminous than full disks amongst K0--M1.75 stars, the probability that the two distributions are drawn from the same parent population is $p=0.13$ and therefore the differences are not significant. Even though the evolved and debris-type systems were not detected, we infer from the upper limits that these disks are 
significantly less luminous than the full and transitional disks with $p=2\times10^{-5}$ and 0.007, respectively.

The right panel in Figure~\ref{fig:grouptype} compares the submillimeter continuum luminosity distributions for stars with spectral types between M4 and M5. In this spectral type range, full disks are comparable in brightness to transitional disks ($p=0.56$). Evolved disks and debris-type disks are significantly fainter than full disks with $p=5\times10^{-4}$ and $4\times10^{-4}$, respectively.  Results are not shown for spectral types between M2 and M3.75 since there is only one transitional disk and 5 evolved disks, and none were detected.

In summary, we find that full and transitional disks have no discernible differences in the submillimeter continuum flux distributions for spectral types between K0 and M5. However, full disks are more luminous in the Band 7 continuum than both evolved disks and debris-type systems.

\section{Analysis of ALMA \coj\/ Data}
\label{gas}

Images of the integrated \coj\/ intensity and the CO spectra are presented in Figure~\ref{fig:images_all} for all sources. In this section, we describe the procedure to measure CO fluxes (Section~\ref{subsec:coflux}) and analyze the results with disk demographics (Section~\ref{subsec:codemo}).

\subsection{Flux measurements}
\label{subsec:coflux}

The spectral setup encompassed the \coj\/ spectral line as a tracer of the gas content of the disks. Measuring robust limits on the CO emission is more challenging than measuring the dust continuum since the stellar velocity is not typically known to high accuracy and the line width can range from a couple to tens of km s$^{-1}$ depending on the inclination angle of the disk with respect to the line of sight, the radial extent of the CO emission, and the mass of the star.

The CO fluxes were measured using aperture photometry, but allowing the aperture radius and velocity range to be tuned to each source. The CO images were deconvolved using clean with a threshold limit of 2$\sigma$, where $\sigma$ is the noise level in the spectral cube. A spectrum was computed solely from the clean components within  a 0.25\arcsec\/ radius centered on the position of the continuum source or the stellar position if the continuum is not detected. If fewer than two channels in the spectrum had intensities less than 3$\sigma$ between velocities of 0 and 10~\kms, then no additional fine tuning of the radius and velocity was performed. If two or more channels had intensities greater than 3$\sigma$, the procedure moved to optimize the velocity range and then the radius. The velocity range was defined as the largest continuous velocity range that contained clean components around the peak of the line. The radius was defined as the radius that contains 95\% of the integrated flux of the cleaned components within a radius of 1\arcsec, twice the FWHM size of the continuum source, or twice the aperture radius, whichever is greater. The aperture radius was manually adjusted for 40 sources based on examination of the results.

The integrated CO flux was computed by summing the channels within the radius and velocity range defined above. The uncertainty in the integrated flux was estimated by randomly placing 100 apertures of the same size and velocity extent within the image cube and computing the standard deviation of the measurements. Table~\ref{tbl:flux_co_sample} summarizes the CO flux measurements for the 284 sources that were targeted by ALMA. Table~\ref{tbl:flux_co_other} presents the CO flux measurements for the additional 40 continuum sources that were detected in the field. The table includes the aperture radius, the velocity extent used to compute the flux, the integrated flux, and the 1$\sigma$ uncertainty. 

\begin{figure}
\centering
\includegraphics[width=\columnwidth, trim=0.7in 0 0.7in 0, clip]{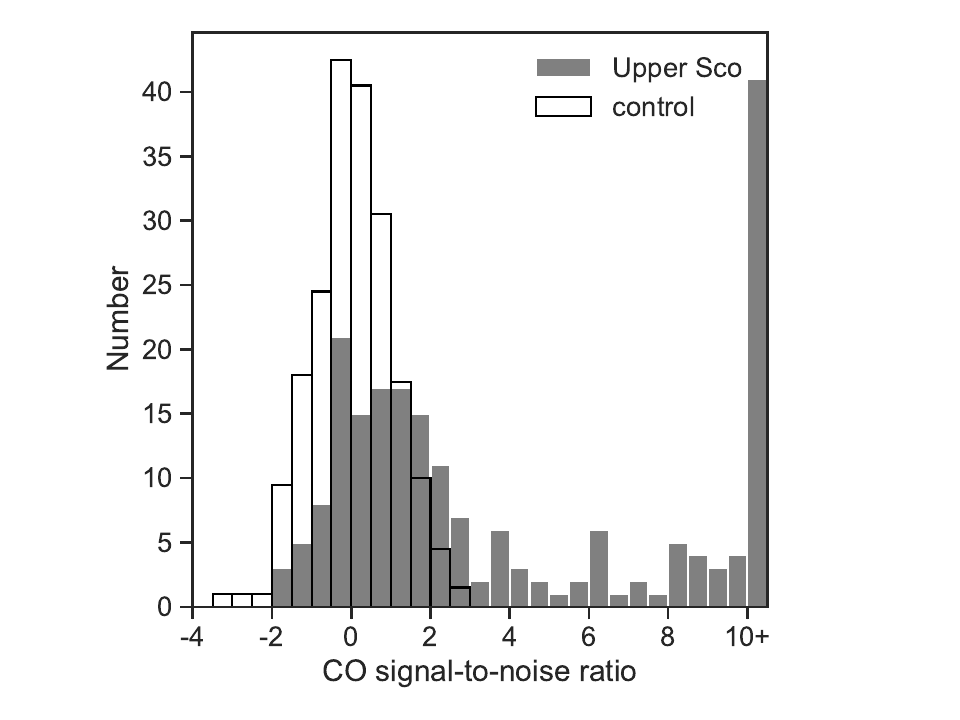}
\caption{Histogram of the signal-to-noise ratio of the CO J=3--2 integrated intensity for the 199 Upper Sco members (gray histogram) and the control sample (open histogram).
\label{fig:co_snr}}
\end{figure}

To estimate the number of false positive detections from random noise, the automated CO identification procedure was applied toward the stellar positions but for velocities between $-$50 and $-$40~\kms\ and between 40 and 50~\kms, where no CO detections are expected. The procedure to identify lines and measuring the \co\/ velocity extent and aperture radius followed that as for actual CO identifications. The open histogram in Figure~\ref{fig:co_snr} shows the distribution of these control velocity ranges, where the positive and negative velocity ranges were weighted by half to form the histogram. The maximum signal to noise in the control sample is 2.6. The Upper Sco sample significantly exceeds the control sample for signal-to-noise ratios $\gtrsim$ 3, suggesting that most of these measurements are real detections.

\subsection{Demographics}
\label{subsec:codemo}

As with the continuum data, we now examine the demographics of the \coj\/ flux measurements by spectral type and disk type. In addition, we compare the \coj\/ and continuum fluxes.

\subsubsection{Spectral type}

\begin{figure}
\centering
\includegraphics[width=\columnwidth]{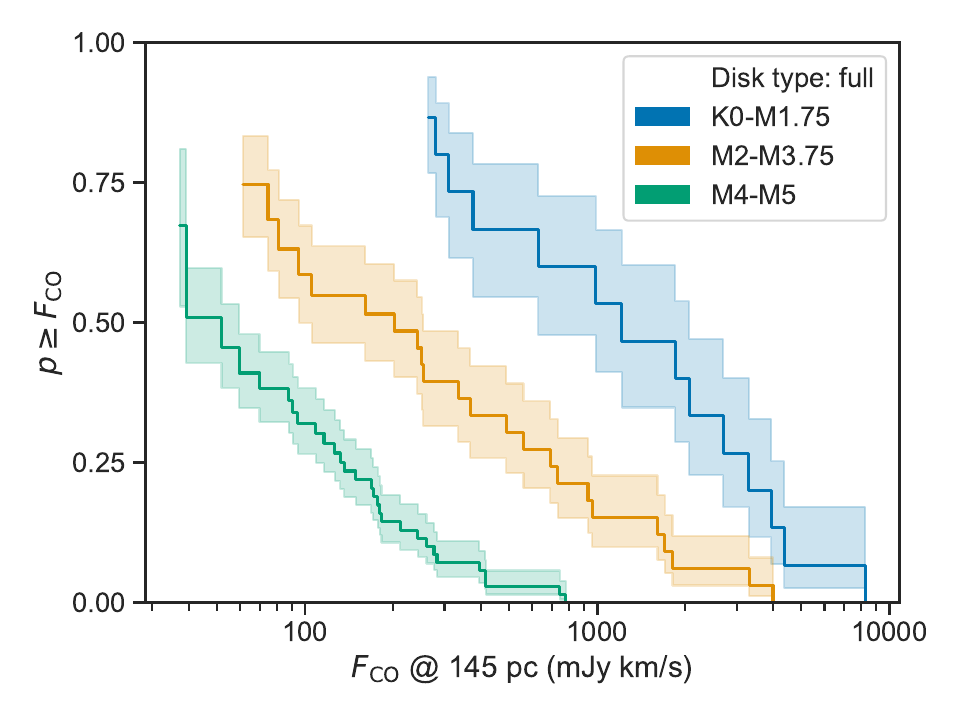}
\caption{The Kaplan-Meier estimator of the \coj\/ fluxes for full disks in Upper Sco, grouped in three spectral type ranges. Fluxes have been scaled to  a distance of 145~pc. The results show that earlier spectral types are more luminous in CO than disks around later spectral types. 
\label{fig:groupspt_co}}
\end{figure}

Figure~\ref{fig:groupspt_co} shows the cumulative distribution of \coj\/ fluxes for the Upper Sco sample for the full disks. The results show a clear dependence of the submillimeter flux distributions with spectral type. Full disks surrounding stars with spectral types K0--M1.75 are clearly brighter than M2--M3.75 stars ($p=2\times10^{-4}$), which in turn are brighter than the disks around M4--M5 stars ($p=10^{-4}$). The median \coj\/ luminosity for disks around K0--M1.75 stars is $\sim6\times$ higher than for M2--M3.75 stars and $\sim23\times$ higher than for M4--M5 stars.

\subsubsection{Disk type}

As in the continuum (see Section~\ref{corr_disktype}), the overall \coj\ detection rate varies with disk type. The detection rates for full and transitional disks are 57\% (69/121) and 73\% (11/15), respectively. By contrast, evolved disks have a detection rate of 7\% (2/28), and only one of the 38 debris-type disks were detected. The one debris-disk type source detected is the Class~III source 2MASS J16052076-1821367. As shown in the figure set for Figure~\ref{fig:images_all}, continuum emission was not detected toward the stellar position, but there is continuum emission and extended CO emission in the region. Therefore, the CO emission toward this source may be from the ambient cloud and not a disk. 

\subsubsection{Comparison with the continuum flux}

\begin{figure}
\centering
\includegraphics[width=\columnwidth]{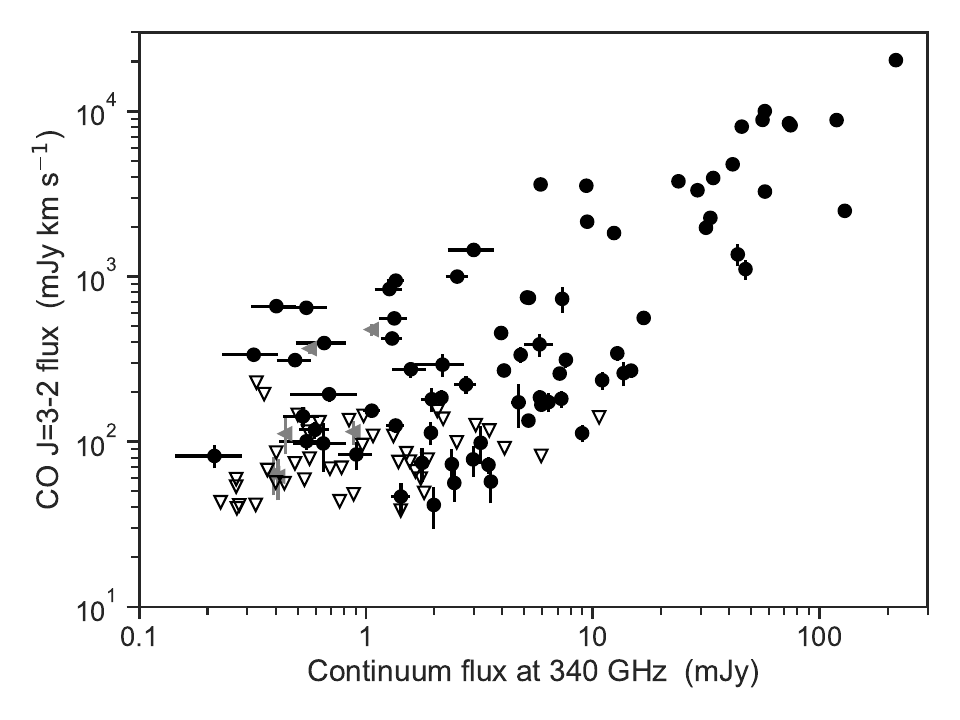}
\caption{Correlation between the \coj\/ integrated intensity and the Band 7 continuum flux for the Upper Sco members. Filled circles indicate 77 sources that are detected in both \coj\/ and the continuum. Gray-filled triangles represent six sources detected in \coj\/ but not the continuum, and open triangles represent 43 sources detected in the continuum but not \coj. Non-detections are plotted as 3$\sigma$ upper limits. 
\label{fig:co_cont}}
\end{figure}

Figure~\ref{fig:co_cont} shows the correlation between the \coj\/ integrated intensity and the Band 7 continuum flux for the Upper Sco members. In general, the brighter continuum disks and are also brighter in \coj.  This general trend is consistent with that previously found in Upper Sco \citep{Barenfeld17} and other regions \citep{Long22,Sanchis21}. If  both the line and continuum emission are optically thick, such a correlation is expected since both tracers are basically tracing the surface area of the disk, although the spatial extent of the gas tends to be larger than the dust, likely due to radial drift of the dust \citep{Sanchis21,Long22}.

\section{Discussion}
\label{discussion}

Given the empirical trends identified in the data, we examine the physical implications and compare the Upper Sco data with other regions.

\subsection{Variation of luminosity with spectral type}

One factor contributing to the difference in submillimeter luminosities with spectral type is that stars with later spectral types have lower stellar luminosities, and thus the dust will tend to be cooler and less luminous. However, the differences in submillimeter luminosities are far larger than can be explained by differences in the stellar luminosities alone. In the Rayleigh-Jeans limit, the submillimeter luminosity will scale linearly with the dust temperature. For optically thin dust emission, a dust grain that emits as a blackbody at radius $R_\mathrm{dust}$ from the star will have a dust temperature that scales as $L_*^{0.25} R_\mathrm{dust}^{-0.5}$. The median submillimeter luminosity around K0--M1.75 stars in Upper Sco is $\sim$ 23 times higher than around M4--M5 stars. Since an early K-type star in Upper Sco is $\sim$ 16 times more luminous than an M5 star \citep{Barenfeld16}, the stellar luminosity differences can only account for a factor of two difference in the dust luminosities. Alternatively, assuming the Band 7 dust emission is optically thick, the dust continuum emission would be proportional to the disk radius squared. Disks around the later M-type stars would need to be $\sim$ 5$\times$ smaller around later type stars to explain the low luminosities after accounting for the variation in the dust temperature with radius. \citet{Barenfeld17} found that the size of the continuum disks in Upper Sco is a factor of $\sim3$ smaller than found in Ophiuchus, Lupus, and Taurus \citep{Tripathi17,Tazzari17}. \citet{Hendler20} conducted a more self-consistent analysis and found the median disk size in Upper Sco is  a factor of 1.7 times smaller than in Lupus and 1.3 times smaller than in Chamaeleon~I. The smaller disk sizes in Upper Sco can account for some but not all of the differences in the dust continuum luminosities. This suggests that in addition to having a more compact size, the disks around low mass stars have reduced dust column densities and masses compared to disks around higher mass stars. This conclusion is analogous to the disk mass vs. stellar mass correlation that has been previously observed in several regions \citep{Andrews13,Pascucci16,Ansdell16,Ansdell17,Tripathi17,Andrews18}. 

\subsection{Comparison to other regions}

Previous studies have shown that the continuum luminosity of disks in Upper Sco tend to be lower compared to most younger regions, including relative to Taurus \citep{Barenfeld16}, Lupus \citep{Ansdell16}, Chamaeleon I \citep{Pascucci16}, Orion \citep{Eisner18}, IC 348 \citep{Ruiz18},  $\sigma$ Orionis \citep{Ansdell17}, and NGC 2024 \citep{vanTerwisga20}. 
$\lambda$ Orionis has disk luminosities comparable to Upper Sco, but it also has an older inferred age than nearby star-forming regions \citep{Ansdell20}.  
\citet{Cazzoletti19} found that disks around stars near the Corona Australis cloud have comparable masses as disks in Upper Sco even though the embedded population is much younger \citep[$<$3~Myr][]{Meyer09,Sicilia11}. However, subsequent work has shown that many of those disks reside in an older population surrounding the cloud that is closer to the age of Upper Sco \citep{esp22}. 

The most common interpretation of the comparison between regions is that the disks in Upper Sco are fainter due to evolutionary effects. Upper Sco tends to be older than the comparison regions, and the fainter disk luminosities are presumed to indicate the depletion of millimeter-sized dust grains in disks with increasing age. However, other factors are known to influence disk evolution, including proximity to massive stars \citep{Ansdell17} and stellar multiplicity \citep{Harris12,Akeson19}. \citet{Trapman20b} showed that the sustained higher radiation fields in OB associations could also impact the masses of disks, which could contribute to the luminosity differences in addition to stellar age.

A more refined comparison of the disk luminosities is warranted given the extended ALMA sample. Because of the dependencies on spectral type (see Section~\ref{corr_spt} and disk type (see Section~\ref{corr_disktype}), it is important to select stars with similar properties. Accordingly, we compared Upper Sco with submillimeter surveys where spectral types are available and the disks have been classified with the same infrared classifications. These regions include Lupus \citep{Luhman20a}, Ophiuchus \citep{Esplin20}, Taurus \citep{esp19,luh23tau}, and Chamaeleon I \citep{luh08cha1,luh08cha2,luh10tau}

Taurus deserves special mention, since it has been extensively studied at all wavelengths, especially searching for binaries and measuring the submillimeter fluxes for the individual stellar components \citep{Akeson19}. However, the mid-infrared by {\it Spitzer} and {\it WISE} often resolve the individual binary components and the disk classification is for the composite system \citep{esp19}. For cases where the disk classification was for the composite system, we assigned the disk type to the primary and removed the secondary from the sample. In one case (CZ Tau), the primary was not detected in the millimeter continuum and the secondary component was, making it ambiguous if the disk classification (full) is appropriate for the primary, the secondary, or both. This star was also excluded from the sample following what was done for Upper Sco.
 
The observed submillimeter fluxes have been scaled to a common distance of 145~pc using the distances inferred from {\it Gaia} parallaxes. If the distance is not available from {\it Gaia} or the uncertainty on the distance is more than 10~pc, then the median distance of the members in the star forming region was used. The submillimeter fluxes were all scaled to a common frequency assuming the submillimeter spectral energy distribution has a slope of $\alpha=2.2$ \citep{Tazzari20b} for $S_\nu \propto \nu^\alpha$.
 
\begin{figure*}
\centering
\includegraphics[width=0.5\textwidth]{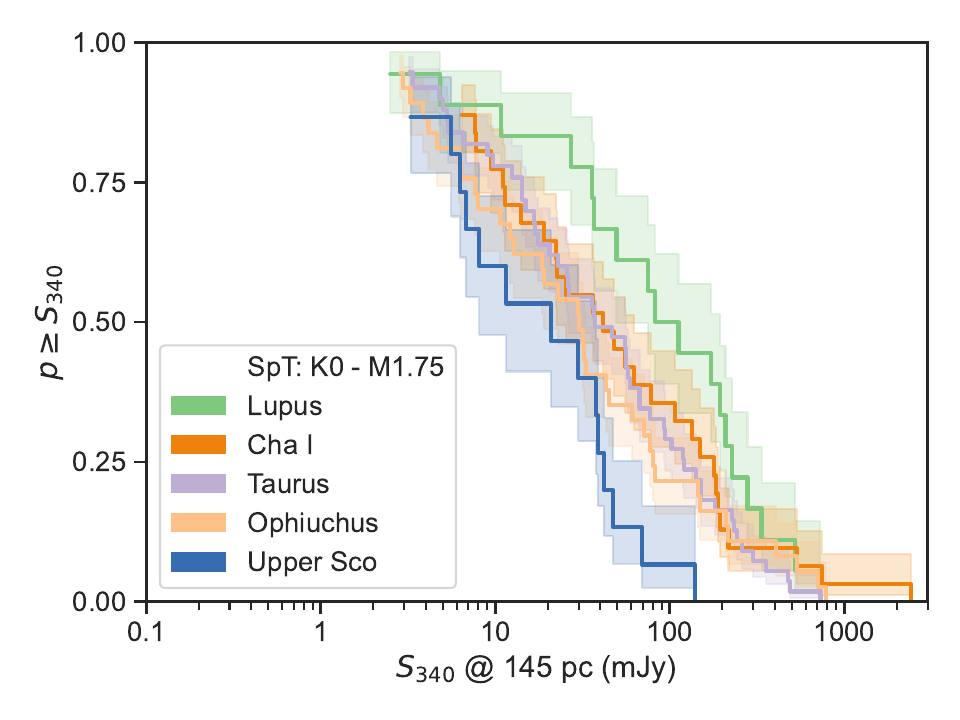}
\includegraphics[width=0.5\textwidth]{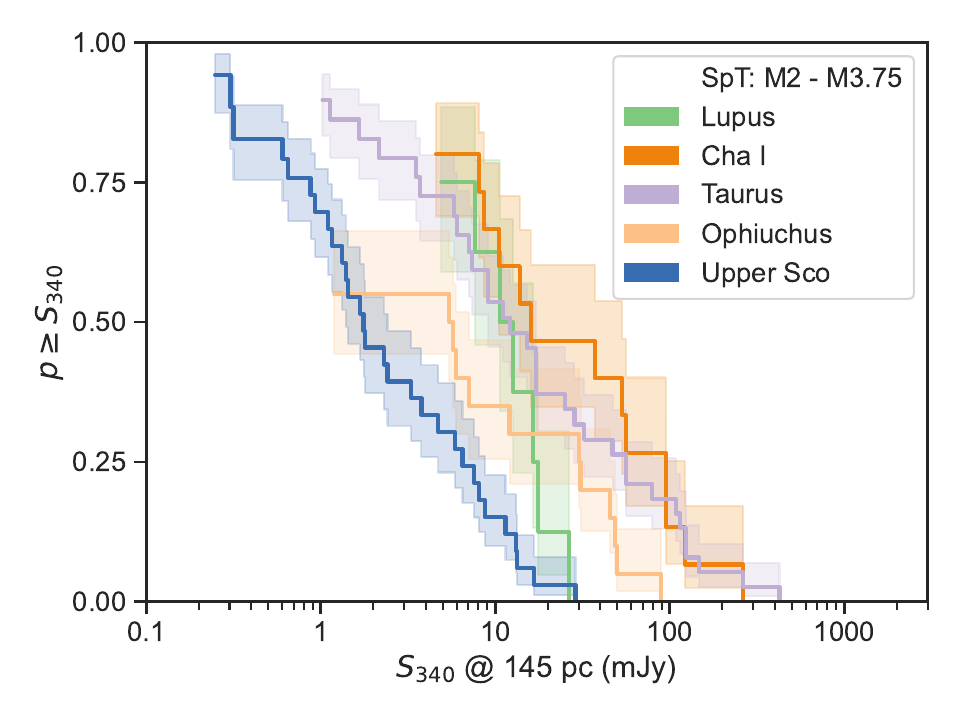}
\includegraphics[width=0.5\textwidth]{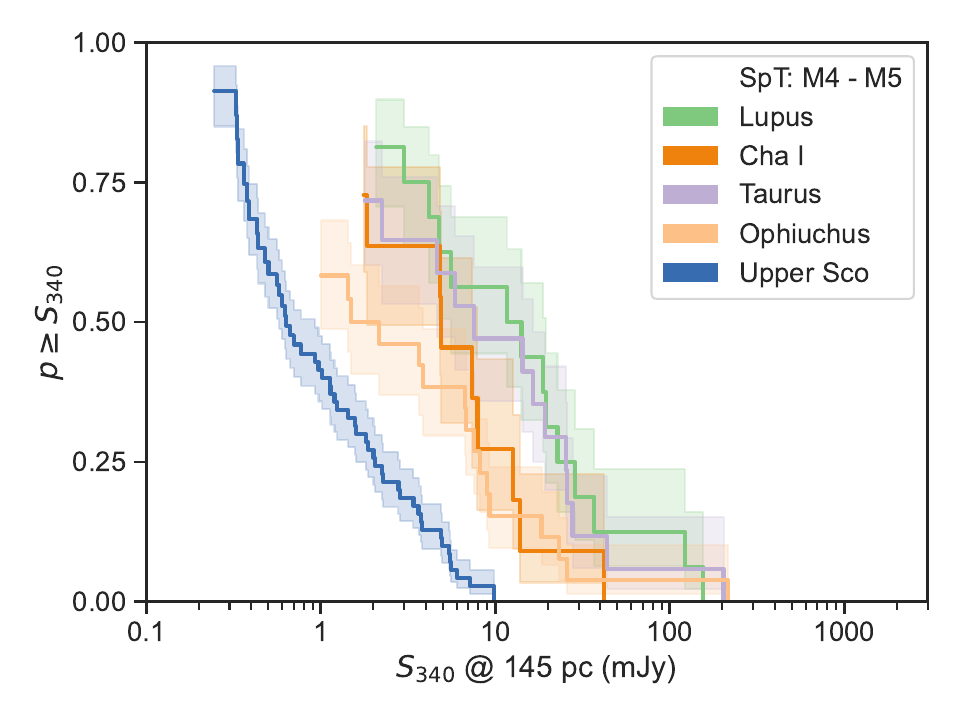}
\caption{The Kaplan-Meier estimator of the submillimeter fluxes at 340~GHz for full disks in Lupus \citep{Ansdell16}, Ophiuchus \citep{Williams19}, Taurus \citep{Akeson19}, Chamaeleon~I \citep{Pascucci16}, and Upper Sco (this paper). Each panel shows the results for different spectral type ranges. Only disks classified as full are shown. All fluxes have been scaled to a distance of 145~pc. 
\label{fig:compare}}
\end{figure*}

Figure~\ref{fig:compare} compares the submillimeter continuum flux distribution in Upper Sco to those in Lupus,  Ophiuchus, Chamaeleon~I, and Taurus, where the comparisons are shown for three different spectral type ranges. The most rigorous comparison would be done by stellar mass instead of spectral type to account for pre-main-sequence evolution between ages of $\sim$ 1 to 10~Myr. However, the change in effective temperature (i.e., spectral type) with stellar age is expected to be small, since pre-main-sequence stars decrease in luminosity with age with small changes in temperature. For example, using the \citet{Baraffe15} evolutionary models, the photospheric temperature of a 0.1, 0.2, and 0.5~M$_\odot$ star (approximately spectral types of M4.5, M3.5, and M0.5, respectively) varies by $\sim$ 110~K, 50~K, and 150~K between ages of 1~Myr and 10~Myr. This corresponds to a change in spectral type of approximately 1 subclass or less \citep{Pecaut13}. Therefore, within a spectral type range, the stellar masses should be approximately the same across regions. 
The comparison shows that for M4--M5 stars, Upper Sco clearly has fainter disk continuum luminosities than Lupus ($p=9\times10^{-9}$), Ophiuchus ($p=0.007$), Chamaeleon~I ($p=7\times10^{-5}$), and Taurus ($p=7\times10^{-6}$). The same trends are present for M2--M3.75 stars, but any differences are significant only for Taurus ($p=3\times10^{-4}$) and Chamaeleon~I (p=$2\times10^{-6}$), marginally for Lupus ($p=0.02$), and not significant for Ophiuchus ($p=0.87$). For the earliest spectral types shown (K0--M1.75), Upper Sco continues to have fainter disk luminosities, but the most significant difference is for Lupus with $p=0.007$. 
 
In summary, the disk continuum luminosities tend to be lower in Upper Sco than in other regions as found in previous analyses. The differences are more pronounced in the later (M4--M5) spectral types than in earlier type stars (K--M1.75). This trend with spectral type is analogous to the results in \citet{Pascucci16}, who found that the disk luminosity-stellar mass relation steepens with stellar age. They suggest this trend can be explained by models incorporating grain growth, fragmentation, and radial drift where, if disks are in the fragmentation regime, the timescale over which dust is depleted is faster in lower mass stars.

\subsection{Clusters in Upper Sco}

\citet{Ratzenbock23_cluster} identified nine clusters within Upper Sco that are co-spatial and co-moving based on the precise astrometry from {\it Gaia}. The ages of clusters inferred by placing the stars in the Hertzsprung-Russell (H-R) diagram range from approximately 4~Myr to 19~Myr \citep{Ratzenbock23_time} when adopting the PARSEC models \citep{Marigo17} and the {\it Gaia} $G$, $G_{\rm BP}$, and $G_\mathrm{RP}$ photometry. 

Of the 202 stars analyzed in this paper as Upper Sco members as defined in \citet{Luhman22}, 200 appear in the analysis by \citet{Ratzenbock23_cluster}, all of which are associated with one of the Upper Sco clusters. In contrast, of the 284 stars presented in this paper, 248 are associated with a cluster in \citet{Ratzenbock23_cluster}, of which 40 are defined as an Upper Sco member in \citet{Ratzenbock23_cluster} but not in \citet{Luhman22}. It is beyond the scope of this paper to analyze the formal definition of Upper Sco, so here we consider the 200 stars that are defined as Upper Sco members by both \citet{Ratzenbock23_cluster} and \citet{Luhman22}.   

To investigate if the continuum properties of the disks vary amongst the clusters defined in \citet{Ratzenbock23_cluster}, we grouped clusters 1 and 2 (age 3.8-5.8~Myr), clusters 3 and 4 (age 7.6-9.8~Myr), and clusters 5 and 6 (age 10-12.7~Myr). Considering full disks only, this resulted in a sample of 30, 75, and 16 stars, respectively. None of the Upper Sco members defined here with disks are members of clusters 7, 8, and 9 in \citet{Ratzenbock23_cluster}. 

\begin{figure}
\centering
\includegraphics[width=0.5\textwidth]{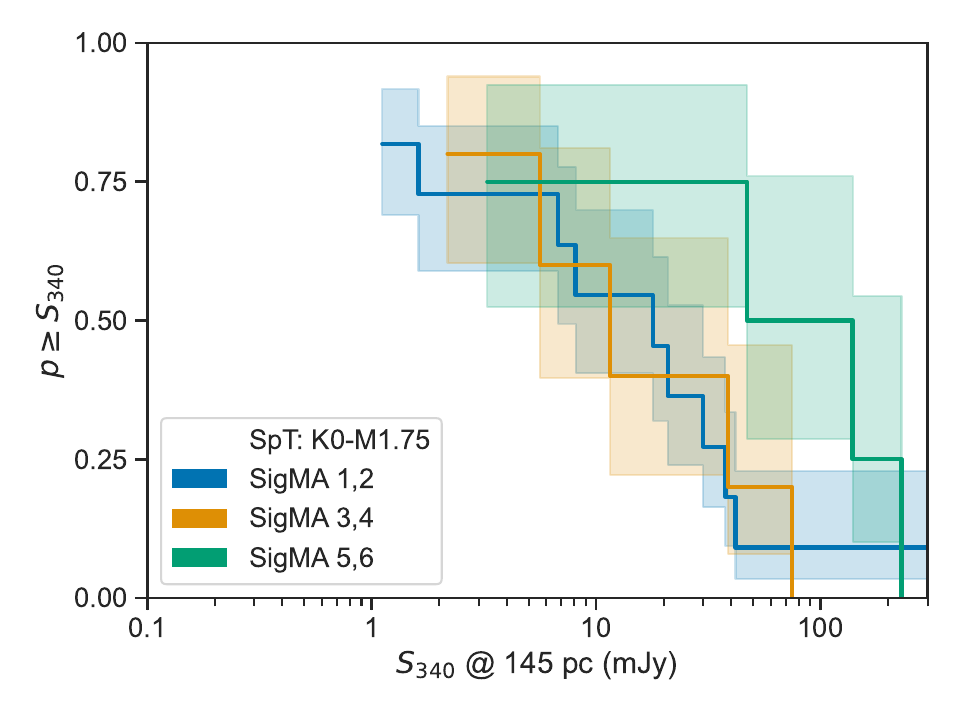}
\includegraphics[width=0.5\textwidth]{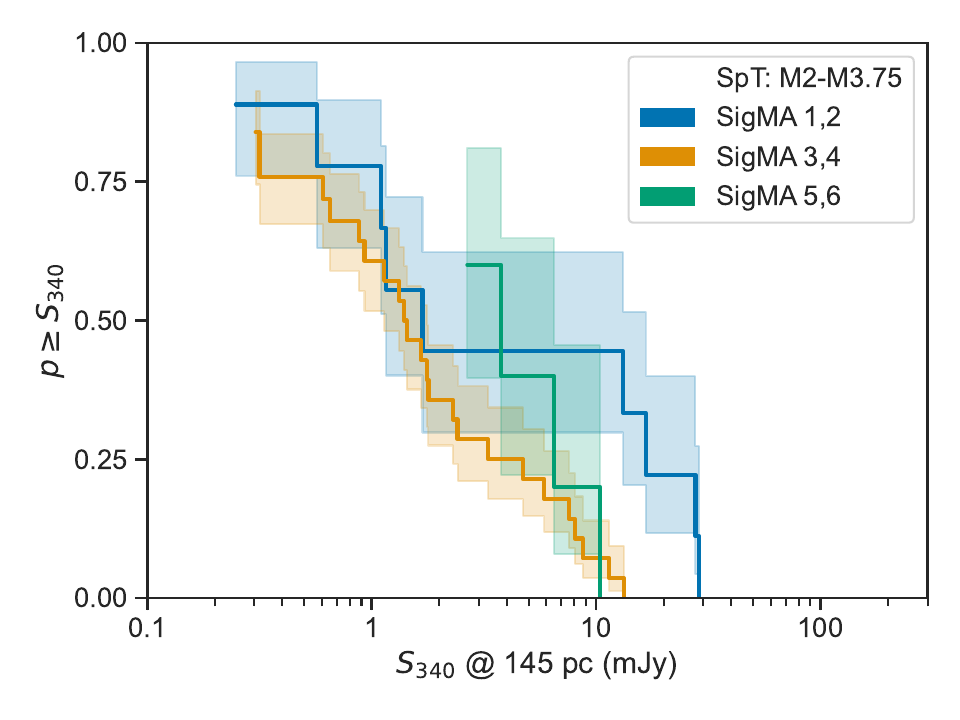}
\includegraphics[width=0.5\textwidth]{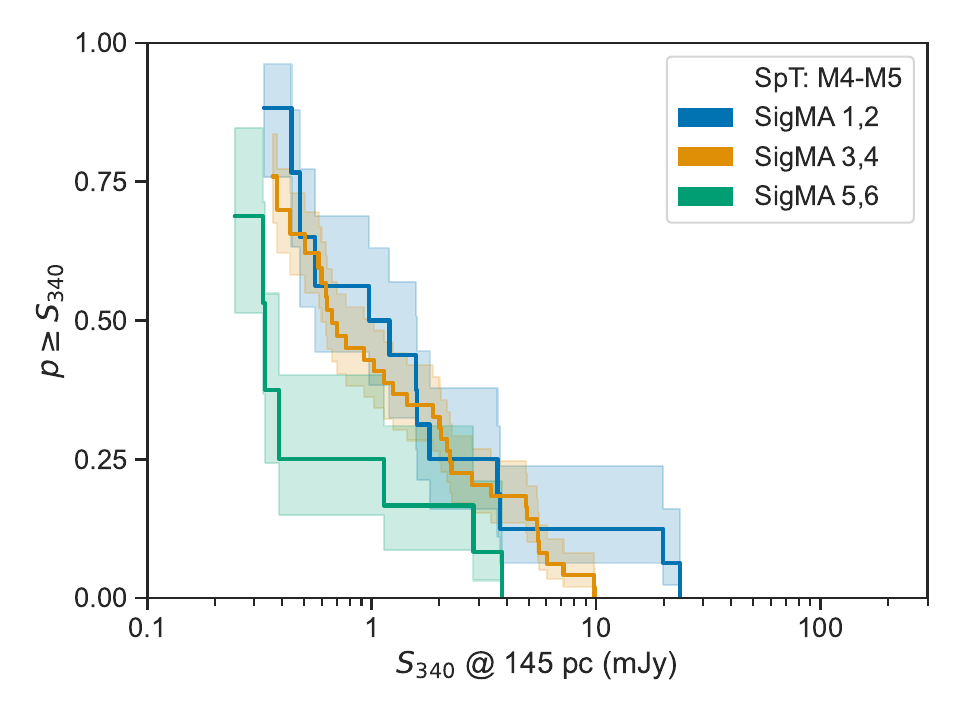}
\caption{The Kaplan-Meier estimator of the Band 7 continuum fluxes for full disks in Upper Sco for stars with spectral type K0 to M1.75 (top panel), M2 to M3.75 (middle panel), and M4 to M5 (bottom panel). Within each panel, distributions are shown for members of the Upper Sco clusters identified by \citet{Ratzenbock23_cluster}.
\label{fig:clusters}}
\end{figure}

Figure~\ref{fig:clusters} shows the cumulative distributions of Band 7 continuum fluxes grouped by spectral type in the panels and by cluster within a panel. For the spectral-type ranges considered here, no significant differences are observed in the flux distributions for the clusters. The most significant difference is for M4--M5 stars in clusters 5--6, which tend to have less luminous disks than clusters 1--2 ($p=0.009$) and 3--4 ($p=0.04$). Although this is consistent with less luminous disks at older ages, the probabilities are only marginally significant. In addition, there are no obvious differences in the flux distributions for the other spectral types. Any differences in the dust continuum luminosity distributions between the clusters are evidently less than those observed between  regions (Figure~\ref{fig:compare}). Therefore, if there is any disk evolution between the clusters, it is not readily apparent in the submillimeter continuum fluxes of the surviving disks.

\section{Summary}

The Upper Sco OB association is one of the benchmark regions for measuring the evolution of protoplanetary disks. It represents the largest collection of $\sim$ 10~Myr stars in the solar neighborhood and it is at comparable distances to the young (1~Myr) star forming regions. Since the initial ALMA studies of disks in Upper Sco \citep{Carpenter14,Barenfeld16,Barenfeld17}, the census of the disk population of the association has more than doubled through a combination of ground-based spectroscopy, {\it Gaia} astrometry, and mid-IR photometry from {\it WISE} \citep{Esplin18,Luhman20b,luh22disks}.
In this study, we have presented ALMA data for newly identified disks in Upper Sco. The observations were obtained in the Band 7 (340~GHz) continuum and in \coj. Combined with previous observations, the full sample contains 284 stars. Of these stars, 208 are now considered Upper Sco members based on {\it Gaia} kinematics, and 202 form the sample studied here after removing possible contaminants (see Appendix~\ref{appendix:contaminants}). Of the 202 Upper Sco members analyzed here, 120 have been detected in the Band 7 continuum and 83 in \coj\/ with a signal-to-noise ratio of $\geq3$.

The expanded sample reinforces the conclusions of previous studies. For full disks, the submillimeter luminosities show a trend with stellar spectral type in that disks around later-type stars are fainter the submillimeter continuum than disks around earlier type stars for spectral types between K0 and M5. The magnitude of the difference cannot be explained by the differences in stellar luminosities or the differences in the typical disk size \citep{Barenfeld17,Hendler20}, and may indicate reduces dust masses for the later type stars as found in previous studies \citep{Andrews13,Pascucci16,Ansdell16}. The \coj\/ integrated intensities also show a clear variations with spectral type, as the median CO flux around K0--M1.75 stars is $\sim23\times$ brighter than M4-M5 stars. 

The dust luminosities in Upper Sco were compared with nearby star forming regions. As found in previous studies (see review by \citealt{Manara23}), stars with full disks tend to have fainter submillimeter luminosities in Upper Sco compared to younger regions. The difference is most pronounced for later spectral types (M4--M5) than earlier spectral types (K0--M1.75).

\begin{acknowledgments}

We thank the referee for carefully reviewing the manuscript.
This paper makes use of the following ALMA data: ADS/JAO.ALMA\#2011.0.00526.S, ADS/JAO.ALMA\#2013.1,00395.S, and ADS/JAO.ALMA\#2018.1.00564.S. 
ALMA is a partnership of ESO (representing its member states), NSF (USA) and NINS (Japan), together with NRC (Canada), NSTC and ASIAA (Taiwan), and KASI (Republic of Korea), in cooperation with the Republic of Chile. The Joint ALMA Observatory is operated by ESO, AUI/NRAO and NAOJ. 
The National Radio Astronomy Observatory is a facility of the National Science Foundation operated under cooperative agreement by Associated Universities, Inc. 
This research has made use of the VizieR catalog access tool, CDS, Strasbourg, France (DOI: 10.26093/cds/vizier).
Part of this research was carried out at the Jet Propulsion Laboratory, California Institute of Technology, under a contract with the National Aeronautics and Space Administration (80NM0018D0004).

\end{acknowledgments}

\facility{ALMA}

\software{
{\tt Astropy} \citep{astropy:2013, astropy:2018},
{\tt CASA} \citep{McMullin07},
{\tt corner} \citep{corner},
{\tt emcee} \citep{Foreman13},
{\tt EnvStats} \citep{EnvStats},
{\tt Lifelines} \citep{Lifelines},
{\tt R} \citep{R}
}

\clearpage
\bibliography{references}

\appendix

\section{Corner plots for the continuum visibility Gaussian fits}
\label{app:corner}

\begin{figure}
\centering
\includegraphics[width=\columnwidth]{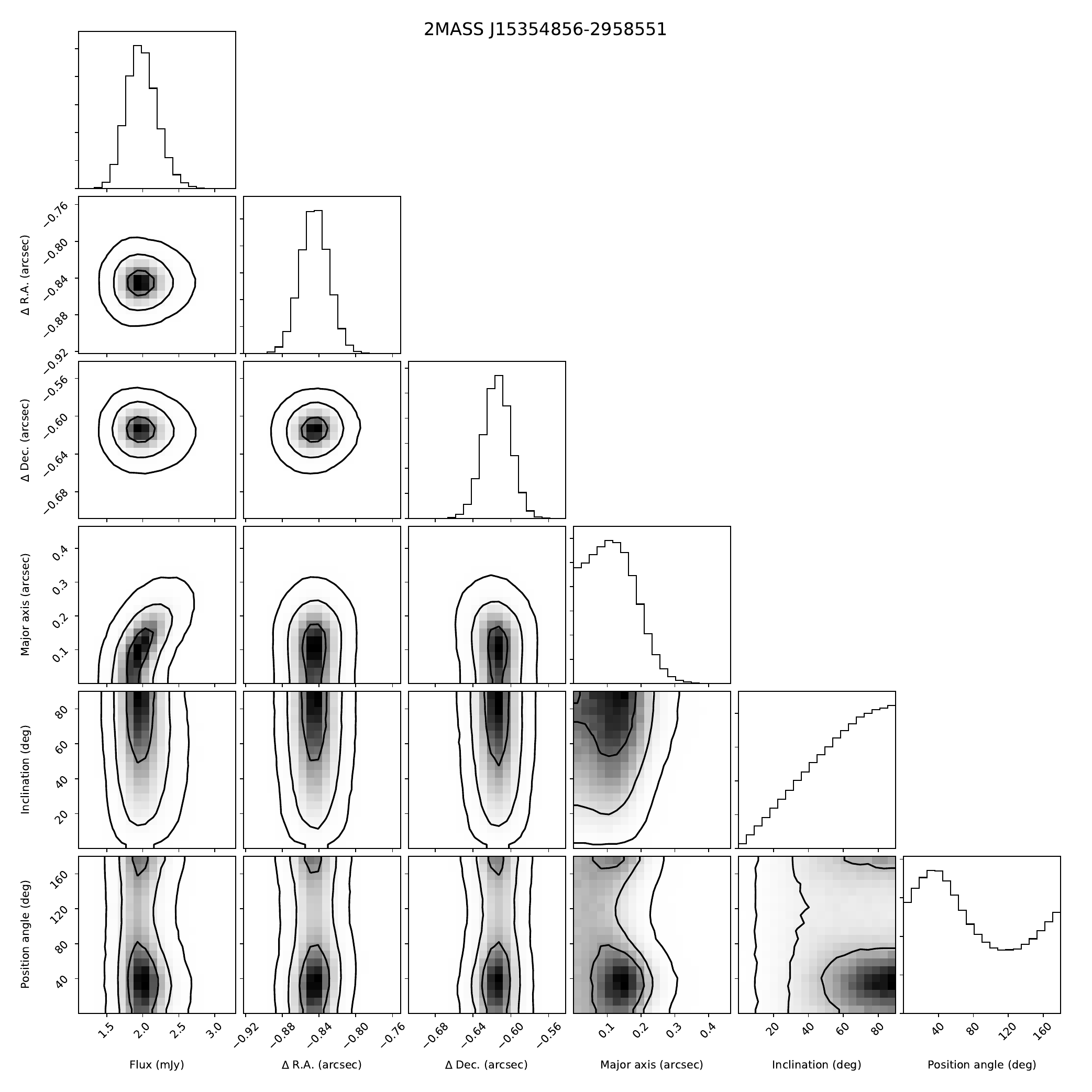}
\caption{Corner plot for the Gaussian fit to the continuum visibilities of 2MASS J15354856-2958551. The diagonal panels show 1-D histograms for each model parameter, marginalized over the others, with dashed vertical lines indicating the best-fit values. The off-diagonal panels present 2-D projections of the posterior probability distributions for parameter pairs, with contours corresponding to the 1$\sigma$, 2$\sigma$, and 3$\sigma$ confidence regions. The full set of figures for the Gaussian fits (121 images) is available in the online journal.
}
\label{fig:corner}
\end{figure}

Figure~\ref{fig:corner} presents the corner plot \citep{corner} that summarizes the parameters from the Gaussian fit to the continuum visibilities for 2MASS J15354856-2958551. Similar plots for sources that were fitted with a Gaussian are shown in the figure set. In cases where multiple source components were fitted to the visibilities, only the parameters associated with the intended stellar target are shown. 

\figsetstart
\figsetnum{13}
\figsettitle{Corner plots for the Gaussian fits to the continuum visibilities}

\figsetend

\section{Identification of the ALMA sources}
\label{appendix:contaminants}

This section analyzes the ALMA continuum sources to identify which sources are coincident with the targeted star and which may be contaminants, either from galaxies,  binary companions, or other members of Upper Sco. The stellar astrometry is derived from the Gaia EDR3 and include corrections for proper motion to the epoch of the ALMA observation. Twenty-four sources do not have a proper motion listed in {\it Gaia} EDR3. For these sources, the median proper motion of the sample was used according to the defined group; i.e., for Upper Sco members, the median proper motion of all Upper Sco members was used and for non-Upper Sco members, the median proper motion of non-Upper Sco members was used.

In addition to the {\it Gaia} sources, ALMA sources were matched with sources detected from optical and infrared multiplicity surveys of Upper Sco. Table~5 in \citet{Tokovinin20} summarized the binary companions in Upper Sco based on their own observations and from \citet{Kohler00}, \citet{Kouwenhoven05}, \citet{Metchev09}, \citet{Kraus12b}, and \citet{Lafreniere14}. ALMA sources were also matched with the multiplicity survey of \citet[see also \citealt{Garufi20}]{Barenfeld19}.

\begin{figure}
\centering
\plotone{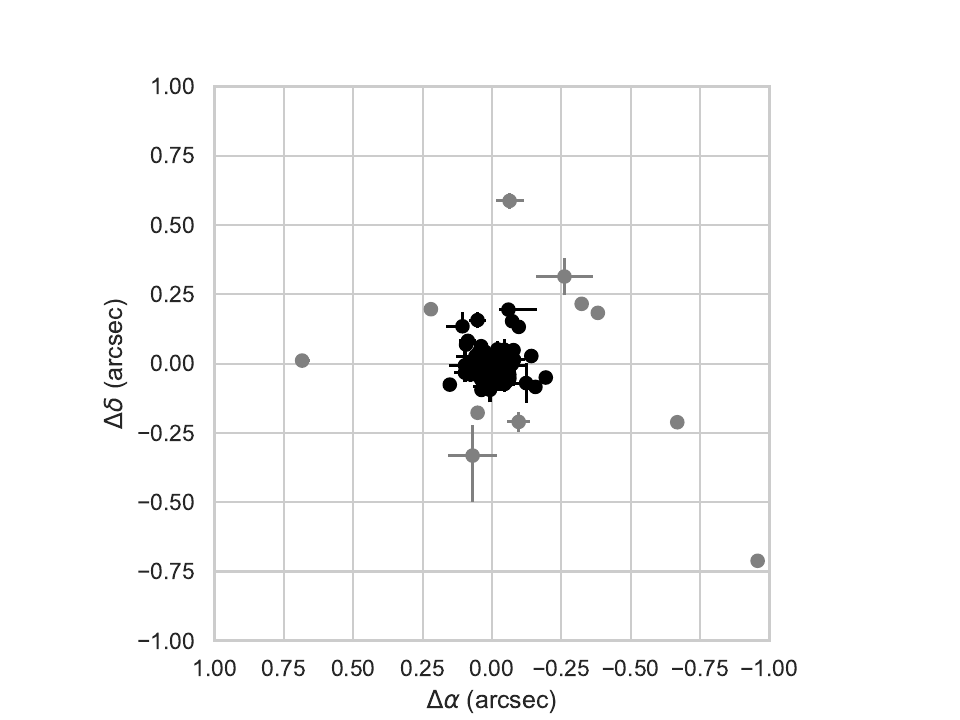}
\caption{Angular separation between the ALMA continuum detection and the stellar position for ALMA sources detected with a signal to noise ratio of $\geq3$. Only sources detected with a $1''\times1''$ region are shown for clarity. If multiple ALMA sources were detected in the image, then only the  ALMA source closest to the star is shown. Black symbols indicate the ALMA  continuum sources that are assumed to be associated with the star, and the gray symbols indicate ALMA sources that are not assumed associated with the star. 
\label{fig:astrometry}}
\end{figure}

Figure~\ref{fig:astrometry} shows the offsets of the ALMA continuum detections (defined as sources with a signal-to-noise ratio of at least 3) to the stellar position within $\pm1''$. Of the 150 sources represented in this figure, 93\% have a separation between the ALMA source and the expected stellar position of $\leq$0\farcs2, with a mean of 0\farcs052 and a dispersion of 0\farcs042. We assume for these sources that the ALMA continuum source is associated with the star with the exception of 2MASS J15442550-2126408, where the ALMA continuum source is more closely aligned with a binary \citep{Tokovinin20}. For the ALMA continuum detections that are more than 0\farcs2 in separation from the expected stellar position, we assume the ALMA source is not associated with the star, as it may be an extragalactic contaminant, a binary, or other member of Upper Sco.

We can estimate contamination based on the source detection statistics in the ALMA images. A total of 25 sources were detected with a flux density $\ge$ 1~mJy between 0\farcs35 and 5\arcsec\ of the stellar position. Such sources could be detected in 231 of the images at a signal-to-noise ratio of at least 5. Assuming conservatively that these are all extragalactic sources, we expect to detect $\sim0.15$ galaxies brighter than 1~mJy within 0\farcs35 of the stellar position. We therefore conclude that the extragalactic contamination is negligible, but some of the contaminants could still be from close binary companions.

After establishing which ALMA sources are associated with the intended stellar target, there are 38 ALMA continuum sources detected with a signal to noise ratio $\geq3$ that are offset from the target star. For 16 of these ALMA sources, the intended stellar counterpart was also detected such that more than one ALMA source appears in the field. This creates ambiguity in the disk classification, since both sources  likely contribute to the infrared emission. Nonetheless, we retain these 16 stars in the sample and keep the disk classification identified in the literature. 

For 22 sources, the targeted star is not detected with ALMA but there is another submillimeter continuum source in the field. 
The infrared excesses for most of these sources were identified from {\it Spitzer} or {\it WISE} with an angular resolution of $\sim6''$. We therefore assume for these targets that when an ALMA source is within 3$''$ of the stellar position, the disk classification is compromised and these sources have been removed from the sample. 

In summary, 13 sources were removed from the sample since the disk classification of the star may be contaminated by another source in close proximity. This includes 7 sources classified as full disks 
(2MASS J15442550-2126408,
2MASS J15564244-2039339,
2MASS J15594231-2945495,
2MASS J16044876-1748393,
2MASS J16050647-1734020,
2MASS J16153220-2010236, and
2MASS J16185382-2053182)
and 6 sources with evolved disks
(2MASS J15524851-2845369,
2MASS J15594460-2155250,
2MASS J16052556-2035397,
2MASS J16064385-1908056,
2MASS J16120505-2043404, and
2MASS J16185277-2259537).
Table~\ref{tbl:flux_other} lists the identified counterparts to the additional sources.

\clearpage
\startlongtable
\begin{deluxetable*}{lllllcc}
\label{tbl:sample}
\tablecaption{Observed sample}
\tablehead{
   \colhead{2MASS} &
   \colhead{Gaia EDR3} &
   \colhead{SpT\tablenotemark{a}} &
   \colhead{Disk Type\tablenotemark{b}} &
   \colhead{Distance\tablenotemark{c}} &
   \colhead{Population\tablenotemark{d}} &
   \colhead{Sample?\tablenotemark{e}}\\
   \colhead{} &
   \colhead{} &
   \colhead{} &
   \colhead{} &
   \colhead{(pc)} &
   \colhead{} &
   \colhead{}
}
\startdata
J15354856-2958551         &  6209142346187345024 & M4         & full                 & $94.1^{+3.2}_{-2.8}$ & c & \\
J15442550-2126408         &  6241506333509857664 & M3.25      & full                 &  &  & \\
J15462696-2443225         &  6236718475767921664 & M4.75      & transitional         & $146.2^{+1.4}_{-1.2}$ & c & \\
J15465432-2556520         &  6234895897806114560 & M5         & full                 & $149.0^{+1.7}_{-1.6}$ & uc & Y\\
J15472572-2609185         &  6234842567193557888 & M4.75      & full                 & $146.0^{+3.4}_{-3.2}$ & c & \\
J15480853-2507011         &  6236644533607453440 & M3         & full                 & $141.1^{+2.1}_{-1.8}$ & c & \\
J15482445-2235495         &  6240396032927158272 & M3.75      & full                 & $140.0^{+0.7}_{-0.7}$ & uc & Y\\
J15485435-2443101         &  6236682741640391808 & M4.75      & full                 & $140.9^{+1.0}_{-0.8}$ & uc & Y\\
\enddata
\tablenotetext{a}{Spectral types are from \citet{Luhman22} and references therein, and from \citet{Manara20}.}
\tablenotetext{b}{Disk types are from \citet{Luhman20a} and \citet{Luhman22}.}
\tablenotetext{c}{Distances are from the Gaia EDR3 distance catalog by \citet{BailerJones21}.}
\tablenotetext{d}{Population subgroup membership in the Scorpius-Centaurus Association from \citet{Luhman20b} and \citet{Luhman22}.}
\tablenotetext{e}{Sources marked with 'Y' are include in the final sample of Upper Sco sources.}
\end{deluxetable*}

\newpage
\begin{deluxetable*}{lcccccc}
\label{tbl:obs}
\tablecaption{Cycle 6 Observations}
\tablehead{
   \colhead{UT date} &
   \colhead{Number} &
   \colhead{Baseline range} &
   \colhead{PWV} &
   \multicolumn{3}{c}{Calibrators}\\
   \cline{5-7}
   \colhead{} &
   \colhead{Antennas} &
   \colhead{(m)} &
   \colhead{(mm)} &
   \colhead{Flux} &
   \colhead{Passband} &
   \colhead{Gain}
}
\startdata
2018 Nov 13 & 43 & 15-1397 & 0.45 & J1427-4206 & J1427-4206 & J1553-2422\\
2018 Nov 15 & 42 & 15-1397 & 0.69 & J1427-4206 & J1427-4206 & J1553-2422\\
2018 Nov 25 & 44 & 15-1397 & 0.84 & J1427-4206 & J1427-4206 & J1553-2422\\
2018 Dec  5 & 44 & 15-783  & 0.37 & J1517-2422 & J1517-2422 & J1553-2422\\
2019 Apr 19 & 49 & 15-740  & 0.64 & J1517-2422 & J1517-2422 & J1551-1755\\
2019 Apr 19 & 48 & 15-740  & 0.60 & J1517-2422 & J1517-2422 & J1626-2951\\
2019 Apr 19 & 48 & 15-740  & 0.57 & J1517-2422 & J1517-2422 & J1626-2951\\
2019 Apr 20 & 46 & 15-740  & 0.43 & J1517-2422 & J1517-2422 & J1553-2422\\
2019 Apr 21 & 46 & 15-740  & 0.74 & J1517-2422 & J1517-2422 & J1553-2422\\
2019 Apr 21 & 47 & 15-740  & 0.82 & J1517-2422 & J1517-2422 & J1553-2422\\
2019 Sep 28 & 50 & 15-2617 & 0.82 & J1517-2422 & J1517-2422 & J1650-2943\\
\enddata
\end{deluxetable*}

\clearpage
\startlongtable
\begin{deluxetable*}{lllllll}
\label{tbl:flux_sample}
\tablecaption{Band 7 continuum flux measurements for the full sample}
\tablehead{
   \colhead{2MASS} &
   \colhead{Flux} &
   \colhead{$\Delta\alpha$} &
   \colhead{$\Delta\delta$} &
   \colhead{Major axis} &
   \colhead{Inclination} &
   \colhead{PA}\\
   \colhead{} &
   \colhead{(mJy)} &
   \colhead{(arcsec)} &
   \colhead{(arcsec)} &
   \colhead{(arcsec)} &
   \colhead{(deg)} &
   \colhead{(deg)}
}
\startdata
J15354856-2958551 & $2.01^{+0.21}_{-0.21}$ & $-0.07^{+0.02}_{-0.02}$ & $-0.00^{+0.02}_{-0.01}$ & $0.10^{+0.07}_{-0.07}$ & $83.5^{+5.5}_{-35.7}$ & $35.5^{+41.1}_{-60.5}$\\
J15442550-2126408 & $-0.23^{+0.15}_{-0.45}$ & \nodata & \nodata & \nodata & \nodata & \nodata\\
J15462696-2443225 & $-0.10^{+0.12}_{-0.12}$ & \nodata & \nodata & \nodata & \nodata & \nodata\\
J15465432-2556520 & $4.71^{+0.35}_{-0.35}$ & $0.07^{+0.02}_{-0.02}$ & $-0.01^{+0.02}_{-0.02}$ & $0.41^{+0.05}_{-0.05}$ & $47.0^{+9.4}_{-13.6}$ & $126.6^{+21.2}_{-17.9}$\\
J15472572-2609185 & $0.23^{+0.11}_{-0.11}$ & \nodata & \nodata & \nodata & \nodata & \nodata\\
J15480853-2507011 & $1.76^{+0.38}_{-0.38}$ & $0.01^{+0.06}_{-0.06}$ & $-0.08^{+0.05}_{-0.05}$ & $0.51^{+0.15}_{-0.11}$ & $71.4^{+13.4}_{-22.1}$ & $46.5^{+28.8}_{-13.5}$\\
J15482445-2235495 & $1.93^{+0.13}_{-0.13}$ & $-0.02^{+0.01}_{-0.01}$ & $0.01^{+0.01}_{-0.01}$ & $0.22^{+0.04}_{-0.04}$ & $86.0^{+3.8}_{-17.7}$ & $167.3^{+16.1}_{-9.1}$\\
J15485435-2443101 & $2.18^{+0.54}_{-0.54}$ & $-0.10^{+0.10}_{-0.09}$ & $-0.06^{+0.06}_{-0.08}$ & $0.83^{+0.28}_{-0.21}$ & $60.3^{+11.9}_{-23.7}$ & $107.1^{+20.4}_{-44.6}$\\
\enddata
\end{deluxetable*}

\clearpage
\startlongtable
\begin{longrotatetable}
\begin{deluxetable*}{llllllll}
\label{tbl:flux_other}
\tablecaption{Band 7 continuum flux measurements for additional sources}
\tablehead{
   \colhead{Field} &
   \colhead{Flux} &
   \colhead{$\Delta\alpha$} &
   \colhead{$\Delta\delta$} &
   \colhead{Major axis} &
   \colhead{Inclination} &
   \colhead{PA} &
   \colhead{Notes}\\
   \colhead{} &
   \colhead{(mJy)} &
   \colhead{(arcsec)} &
   \colhead{(arcsec)} &
   \colhead{(arcsec)} &
   \colhead{(deg)} &
   \colhead{(deg)} &
   \colhead{}
}
\startdata
J15442550-2126408 & $2.92^{+0.50}_{-0.17}$ & $0.05^{+0.01}_{-0.01}$ & $-0.18^{+0.02}_{-0.01}$ & $0.19^{+0.04}_{-0.04}$ & $57.4^{+12.4}_{-19.3}$ & $175.2^{+29.9}_{-12.1}$ & \\
J15480853-2507011 & $1.55^{+0.41}_{-0.41}$ & $-0.03^{+0.06}_{-0.06}$ & $-1.03^{+0.08}_{-0.08}$ & $0.51^{+0.24}_{-0.16}$ & $67.5^{+14.0}_{-23.2}$ & $154.4^{+32.5}_{-28.8}$ & Gaia EDR3 6236644533611109376\\
J15510126-2523100 & $2.41^{+0.24}_{-0.24}$ & $-0.28^{+0.01}_{-0.01}$ & $-2.00^{+0.02}_{-0.02}$ & $0.35^{+0.06}_{-0.05}$ & $83.6^{+5.9}_{-6.0}$ & $1.2^{+2.1}_{-15.4}$ & Gaia EDR3 6236390271550062336\\
J15524851-2845369 & $1.53^{+0.46}_{-0.46}$ & $-0.24^{+0.09}_{-0.11}$ & $0.32^{+0.06}_{-0.07}$ & $0.61^{+0.39}_{-0.24}$ & $61.4^{+18.8}_{-23.1}$ & $80.1^{+52.3}_{-30.1}$ & \\
J15564244-2039339 & $4.24^{+0.13}_{-0.13}$ & $0.86^{+0.01}_{-0.01}$ & $1.05^{+0.01}_{-0.01}$ & $0.15^{+0.02}_{-0.01}$ & $30.7^{+13.9}_{-12.8}$ & $59.2^{+55.7}_{-29.9}$ & Gaia EDR3 6246729391696173440; WDS 15567-2040      \\
J15584772-1757595 & $1.73^{+0.29}_{-0.29}$ & $-3.04^{+0.03}_{-0.03}$ & $0.90^{+0.03}_{-0.02}$ & $0.24^{+0.08}_{-0.09}$ & $55.3^{+23.4}_{-22.3}$ & $102.9^{+70.3}_{-35.3}$ & \\
J15594231-2945495 & $2.03^{+0.60}_{-0.60}$ & $0.07^{+0.09}_{-0.09}$ & $-0.34^{+0.12}_{-0.16}$ & $0.98^{+0.34}_{-0.27}$ & $66.1^{+14.2}_{-24.7}$ & $168.7^{+12.2}_{-42.8}$ & \\
J15594460-2155250 & $0.60^{+0.11}_{-0.11}$ & $0.68^{+0.03}_{-0.03}$ & $0.02^{+0.02}_{-0.03}$ & $0.04^{+0.14}_{-0.04}$ & $83.6^{+5.5}_{-34.9}$ & $122.8^{+31.7}_{-86.5}$ & \\
\enddata
\end{deluxetable*}
\end{longrotatetable}

\clearpage
\startlongtable
\begin{deluxetable*}{llll}
\label{tbl:flux_co_sample}
\tablecaption{\coj\/ flux measurements for the full sample}
\tablehead{
   \colhead{2MASS} &
   \colhead{Flux} &
   \colhead{V$_{\rm LSRK}$\tablenotemark{a}} &
   \colhead{radius\tablenotemark{b}}\\
   \colhead{} &
   \colhead{(mJy km s$^{-1}$)} &
   \colhead{(km s$^{-1}$)} &
   \colhead{(arcsec)}
}
\startdata
J15354856-2958551 & $54.7\pm28.6$ & 0.1 - 9.8 & 0.25\\
J15442550-2126408 & $61.6\pm7.8$\tablenotemark{c} & 5.0 - 7.1 & 0.25\tablenotemark{d}\\
J15462696-2443225 & $63.3\pm53.6$ & 0.1 - 9.9 & 0.25\\
J15465432-2556520 & $172.9\pm51.3$ & 0.1 - 9.8 & 0.25\\
J15472572-2609185 & $58.7\pm49.2$ & 0.0 - 10.1 & 0.25\\
J15480853-2507011 & $294.7\pm48.3$ & -0.1 - 10.0 & 0.25\\
J15482445-2235495 & $113.1\pm18.2$ & -3.0 - 10.2 & 0.25\\
J15485435-2443101 & $292.4\pm45.0$ & 0.0 - 10.1 & 0.25\\
\enddata
\tablenotetext{a}{Velocity range used to compute the flux.}
\tablenotetext{b}{Aperture radius to compute the flux.}
\tablenotetext{c}{Flux measurement is contaminated by nearby source.}
\tablenotetext{d}{Radius set by visual inspection.}
\end{deluxetable*}

\clearpage
\startlongtable
\begin{deluxetable*}{llll}
\label{tbl:flux_co_other}
\tablecaption{\coj\/ flux measurements for additional sources}
\tablehead{
   \colhead{Field} &
   \colhead{Flux} &
   \colhead{V$_{\rm LSRK}$\tablenotemark{a}} &
   \colhead{radius\tablenotemark{b}}\\
   \colhead{} &
   \colhead{(mJy km s$^{-1}$)} &
   \colhead{(km s$^{-1}$)} &
   \colhead{(arcsec)}
}
\startdata
J15442550-2126408 & $152.3\pm18.5$ & -3.9 - 7.1 & 0.25\\
J15480853-2507011 & $588.7\pm50.8$ & -0.1 - 7.5 & 0.30\tablenotemark{d}\\
J15510126-2523100 & $138.6\pm43.1$ & 0.0 - 10.1 & 0.25\\
J15524851-2845369 & $32.7\pm49.5$ & -0.1 - 10.0 & 0.25\\
J15564244-2039339 & $288.8\pm26.7$ & -2.1 - 8.1 & 0.35\tablenotemark{d}\\
J15584772-1757595 & $19.1\pm32.2$ & 0.0 - 10.2 & 0.25\\
J15594231-2945495 & $83.5\pm50.8$ & -0.2 - 10.0 & 0.25\\
J15594460-2155250 & $20.3\pm16.8$ & 0.2 - 9.9 & 0.25\\
\enddata
\tablenotetext{a}{Velocity range used to compute the flux.}
\tablenotetext{b}{Aperture radius to compute the flux.}
\tablenotetext{c}{Flux measurement is contaminated by nearby source.}
\tablenotetext{d}{Radius set by visual inspection.}
\end{deluxetable*}

\end{document}